\documentclass[3p]{elsarticle} 

\biboptions{semicolon}

\usepackage[utf8]{inputenc}
\usepackage{amsmath, amssymb}
\usepackage{color}
\usepackage{subcaption}
\usepackage{gensymb}
\usepackage{hyperref}
\usepackage{bm}

\journal{Applied Mathematical Modelling}

\newcommand{\pdiff}[2]{\frac{\partial #1}{\partial #2}}

\newcommand{\eps}{\epsilon}

\newcommand{\ttt}[1]{\texttt{#1}}

\begin{document}

\begin{frontmatter}

\title{An improved air entrainment model for stepped spillways}

\author[1]{Silje Kreken Almeland\corref{cor1}}
\ead{silje.k.almeland@ntnu.no}
\address[1]{Norwegian University of Science and Technology, Department of Civil
  and Environmental Engineering, NO-7491 Trondheim, Norway}

\author[2]{Timofey Mukha}
\ead{timofey@chalmers.se}
\address[2]{Chalmers University of Technology, Department of Mechanics and Maritime Sciences, Hörsalsvägen 7A, SE-412 96 Gothenburg, Sweden}

\author[2]{Rickard E. Bensow}
\ead{rickard.bensow@chalmers.se}


\cortext[cor1]{Corresponding author}

\begin{abstract}
Numerical modelling of flow in stepped spillways is considered, focusing on a
highly economical approach combining interface capturing with explicit modelling
of air entrainment. Simulations are performed on spillways at four different
Froude numbers, with flow parameters selected to match available experimental
data. First, experiments using the model developed by Lopes et
al.~(Int.~J.~Nonlin.~Sci.~Num., 2017) are conducted. An extensive simulation
campaign is used to carefully evaluate the predictive accuracy of the model, the
influence of various model parameters, and sensitivity to grid
resolution. Results reveal that, at least for the case of stepped 
spillways, the number of parameters governing the model can be reduced. A
crucial identified deficiency of the model is its sensitivity to grid
resolution. To improve the performance of the model in this respect,
modifications are proposed for the interface detection algorithm and the
transport equation for the volume fraction of entrained air. Simulations using
the improved model formulation demonstrate better agreement with reference data
for all considered flow conditions. A parameter-free criterion for predicting
the inception point of air entrainment is also tested. Unfortunately, the
accuracy of the considered conventional turbulence models proved to be insufficient
for the criterion to work reliably. 

\end{abstract}


\end{frontmatter}

\section{Introduction}
\label{sec:intro}

Along with a renewed interest in stepped spillways as a flood overflow structure and energy dissipator in hydraulic engineering, attempts at gaining a better physical description of spillway flows have also intensified.
A process that is especially challenging to study by means of both physical and numerical experiments, is the self-aeration of the spillway.
Yet, since large quantities of entrained air lead to higher flow depths, release
energy, and reduce the potential for damage caused by cavitation, accurate
prediction of aeration is crucial for spillway design.
In this work, mathematical modelling and simulation of air entertainment is in focus.
To put the present contribution into context, a brief review of the physics of air entrainement in spillways is given  below, followed by an overview of past attempts of accounting for them in a numerical setting.

Generally, air entrainment is driven by turbulent motion and occurs when the turbulent forces at the free surface overcome the stabilizing effects of surface tension and buoyancy~\cite{ervine1987}.
Applied to spillways, it has since the early work of \citet{straub&anderson} been widely accepted that the onset of self-aeration takes place when the turbulent boundary layer, developed from the crest, reaches the free surface.
This location is commonly referred to as the `inception point'.
Several contributions consider the onset of the aeration in detail \citep{Cheng2014,Zhang2016,Valero2016}, and empirical relations exist for the distance to the inception point from the spillway crest \citep{Boes2003, matos2001, Bung2011}.
\citet{Boes2003} proposed a computable definition of the inception point as the location where the pseudo-bottom air concentration is $0.01$.

Downstream of the inception point, entrainment quickly leads to a complete
distortion of the perceivable air-water interface into a thick layer occupied by
a mixture of the two phases.
Furthermore, experimental data exhibits a non-negligible concentration of air all the way down to the surface of steps.
The work of \citet{Pfister2011} presents a detailed account of the transport mechanism responsible for that.
It is shown that transiently occurring air throughs can penetrate deep enough to hit the edge of the steps.
This leads to brake-up and eventual entrapment of air pockets in the recirculating flow occupying the corners of the steps.

An important property of stepped spillway flow is that it eventually reaches a
state where its average properties no longer alter in the streamwise direction.
The associated distributions of flow variables are referred to as uniform conditions.
The part of the flow preceding this state is called the development region.
Empirical expressions for the extent of the development region have been given
by several authors \citep{Bung2011,Boes2003b}, as
well as relations for the surface height in the different flow regions along the
spillway~\citep{Bung2011,Boes2003,matos2001}. 

When it comes to numerical modelling of the complicated multiphase physics discussed above, one can generally distinguish two approaches.
One is to try to explicitly capture these phenomena using a high-fidelity scale-resolving simulation framework.
This necessitates using very dense computational grids and therefore consuming vast amounts of computational resources.
For this reason, results from such simulations of spillways have not yet been reported in the literature.
However, works on other aerating flows can be found, e.g.~\cite{Mortazavi2016} for the case of the hydraulic jump.

The alternative approach is to introduce an additional model accounting for the entrainment of air.
Different ways of introducing such modelling have been proposed, also varying in the general multiphase simulation methodology into which they are fit.
Efforts within the framework of the two-fluid model (also referred to as Euler-Euler) have been reported in~\citep{hansch2012multi,wardle2013hybrid,Yan2010, ma2011, moraga2008sub}.
In the context of interface-capturing methods, such as Volume of Fluid (VoF), the general idea is to introduce air entrainment as a subgrid model.
Here the study of~\citet{hirt2003modeling} can be distinguished as pioneering.
This model was implemented in Flow3D\textsuperscript{\textregistered} and has been used in publications on stepped spillways \citep{ValeroBung2015,Dong2019}.
\citet{Lopes2017} incorporated the entrained air flux estimator developed in~\cite{ma2011} into a VoF framework by introducing a separate transport equation for the entrained air.
The solver is implemented in OpenFOAM\textsuperscript{\textregistered}, and results from stepped spillway simulations are presented in the article.

In this work, we present further developments of the model by~\citet{Lopes2017}.
To motivate the need for improvements, results from a simulation campaign, in which the model in its current formulation is applied to spillway flow at four different Froude numbers, are presented.
The simulations constituting the campaign vary in the employed grid resolutions and parameters of the model.
Results from the second campaign, in which the model is modified as proposed here, are then presented, demonstrating better accuracy and robustness.
The article is supplemented by a dataset containing all the simulation results.\footnote{10.6084/m9.figshare.12782339}

The reminder of this article is structured as follows.
Section \ref{sec:cfd} describes the computational methods used in the performed computations.
In Section \ref{sec:simcases}, the setup of the stepped spillway simulations is discussed.
Section~\ref{sec:airentmod} presents the air entrainment model developed by~\citet{Lopes2017}.
Section \ref{sec:aifresults} contains results from the simulation campaign in which the model of~\citet{Lopes2017} is used.
Improvements to the air entrainment model are then proposed in Section~\ref{sec:modeldevel}, and corresponding simulation results are provided in Section~\ref{sec:newres}.
Concluding remarks are given in Section~\ref{sec:conclusion}.

\section{Computational fluid dynamics methods}
\label{sec:cfd}

\subsection{Governing equations}
\label{subsec:governeq}
The flow was simulated using the Volume of Fluid (VoF) multiphase modelling technique
\citep{hirt1981}, in which a single set of governing equations is solved for all phases and the location of the interface is identified based on the values of the cell volume fraction of the liquid phase, $\alpha_l$.
Both fluids are considered incompressible and immiscible.
Furthermore, RANS turbulence modelling is adopted, leading to the following set of governing equations.

\begin{align}
  \pdiff{{\rho \overline u}}{t}+ \nabla \cdot \left(\rho \overline u \otimes \overline u \right) & =
  - \nabla \overline p_{\rho gh} - gx \nabla \rho +
  \nabla \left( \mu \left( \nabla \overline u +
  (\nabla \overline u)^T \right) - \rho \overline{u'\otimes u'}\right) +
  f_s \label{eq:ransmom} \\
\label{eq:ranscont}
\nabla \cdot {\overline u} & =  0.
\end{align}

\noindent
Here the overbar denotes the Reynolds average, $\rho$ is the density, $\mu$ is
the dynamic viscosity, $u$ is the velocity vector, $p_{\rho gh}=p-\rho g \cdot x$ is the dynamic pressure, and $f_s$ is the surface tension force.
The latter is approximated using the Continuous Force Model, see~\cite{brackbill1992} and also~\cite{Rusche2002} for a detailed discussion in the context of OpenFOAM\textsuperscript{\textregistered}.
The term $\rho \overline{u'\otimes u'}$ represents the Reynolds stresses, which are to be approximated by the turbulence model.



An algebraic approach to account for the evolution of $\alpha_l$ is adopted,
with the associated transport equation 
originally formulated as 
\begin{equation}
\label{eq:alpha_mules}
    \pdiff{\alpha_l}{t} + \nabla \cdot (\overline u \alpha_l ) +
    \nabla \cdot \left(u^c (1 -\alpha_l)\alpha_l\right) = 0,
\end{equation}
  in OpenFOAM\textsuperscript{\textregistered}'s VoF solvers.
The third term in the equation is artificial and is meant to introduce additional compression of the interface to ensure its sharpness.
However, here the formulation of this term is modified in order to accommodate it into the air entertainment modelling framework.
The details are provided in Section~\ref{sec:airentmod}.
The definition of $u^c$ is nevertheless not altered: It is aligned with the
interface normal, and its magnitude is computed as $C_\alpha |\bar u|$, where $C_\alpha$ is an adjustable constant, here set to 1.

Given $\alpha$, the material properties of the fluids are readily obtained as
\begin{align}
& \rho = \alpha_l\rho_{l} + (1 - \alpha_l)\rho_{air},
& \mu = \alpha_l\mu_{l} + (1 - \alpha_l)\mu_{air}.
\end{align}
The indices $l$ and $air$ are used to refer to the water and air, respectively.

What remains to be discussed is the choice of turbulence model, which for the case of the stepped spillway is far from trivial.
In principle, the model should be able to properly account for the interaction between the turbulent and multiphase structures in order to provide accurate prediction in the aerated region of the flow.
None of the closures that have found widespread use were developed with this goal in mind.
Nevertheless, it is common for conventional two-equation models to be used for aerated flows.
In~\cite{Lopes2017}, which is the work this article largely builds upon, the
$k$-$\omega$ SST model~\citep{Menter1993} is used for stepped spillway simulations.
In~\cite{ma2011}, it is employed in simulations of a plunging jet and a hydraulic jump.
For the latter, many studies also use the $k$-$\eps$ model and its variations, a
comprehensive list of references can be found in Table 5
in~\cite{Viti2018}. \citet{Qian2009} found the realisable $k$-$\eps$ model to be
favourable for stepped spillway flow. 
Based on the above, we consider both the $k$-$\omega$ SST and the realisable $k$-$
\eps$ model~\citep{Shih1995} and test which of them leads to better predictive
accuracy.

It is pointed out in~\citep{Fan2020} that in the implementation of the above (and also other) turbulence models in OpenFOAM\textsuperscript{\textregistered} the viscous diffusion terms are not treated consistently in regions with a non-zero density gradient.
Since in the simulations of the spillway presented here a density gradient is present across the whole aerated part of the flow, this issue can have a significant effect on the results.
The authors of~\citep{Fan2020} also provide alternative implementations, in which the inconsistency is resolved.
Here, we test using both the default and the improved implementations.

\subsection{Numerical methods}
\label{subsec:nummet}

The computations are performed using OpenFOAM\textsuperscript{\textregistered}
version 5, provided by the OpenFOAM Foundation.
This CFD tool is based on cell-centered finite volume discretization, which is de facto the industry standard.
Two custom solvers are used in the study, implementing the air entraiment modelling presented in Sections~\ref{sec:airentmod} and \ref{sec:modeldevel}.
Both of them represent modifications of the solver \texttt{interFoam}, which is distributed with OpenFOAM\textsuperscript{\textregistered}.
This solver implements the VoF methodology discussed in Section~\ref{subsec:governeq}.
The governing equations are solved in segregated manner using a variant of the PISO algorithm~\cite{Issa1986}.

A crucial component of the numerical setup is the selection of the spatial
interpolation and time integration schemes.
Generally, linear interpolation can be used in space except when considering convective fluxes.
In the momentum equation, the latter are interpolated using the \ttt{limitedLinearV} scheme, which is a TVD scheme based on the Sweby limiter.
The limiter is computed based on the direction of most rapidly changing gradient and then applied to all three velocity components.
This improves stability but at a certain expense in terms of accuracy.
For the convection of $\alpha$ in equation~\eqref{eq:alpha_mules}, a TVD scheme
using the SuperBee limiter is used. 
The van Leer limiter was also considered, but SuperBee led to better results on coarser meshes due to being more compressive.
Unfortunately, in a multi-dimensional setting, using a TVD scheme does not guarantee that the values of $\alpha$ will be bounded between 0 and 1.
Therefore, OpenFOAM\textsuperscript{\textregistered} utilizes an additional flux limiting technique, referred to as MULES.
It is based on the Flux Corrected Transport theory developed Zalasak \citep{Zalesak1979}, more details can be found in~\citep{Damian2013}.
The convective fluxes in the turbulence equations are discretized using the second order upwind scheme, called \ttt{linearUpwind}.
This scheme is unbounded, but no significant effect of parasitic oscillations was observed even on coarse grids.
Finally, as discussed in Section~\ref{subsec:alphag} below, the air entrainment
model adds an advection-diffusion equation for the flow variable $\alpha_g$,
  meant to indicate the distribution of the volume fraction of entrained air
, to the system.
Here, a TVD scheme using the van Leer limiter is employed.

The first-order implicit Euler scheme was used for time-stepping.
The choice is not of particular importance here because the flow eventually arrives to an essentially steady state.
Nevertheless, a CFL number~$\leq 1$ was neccessay to maintain in order to keep the simulations stable.
This was achieved using adaptive time-stepping.

\section{Simulation cases}
\label{sec:simcases}
\begin{figure}[htp]
   \centering
        \includegraphics[width=\textwidth]{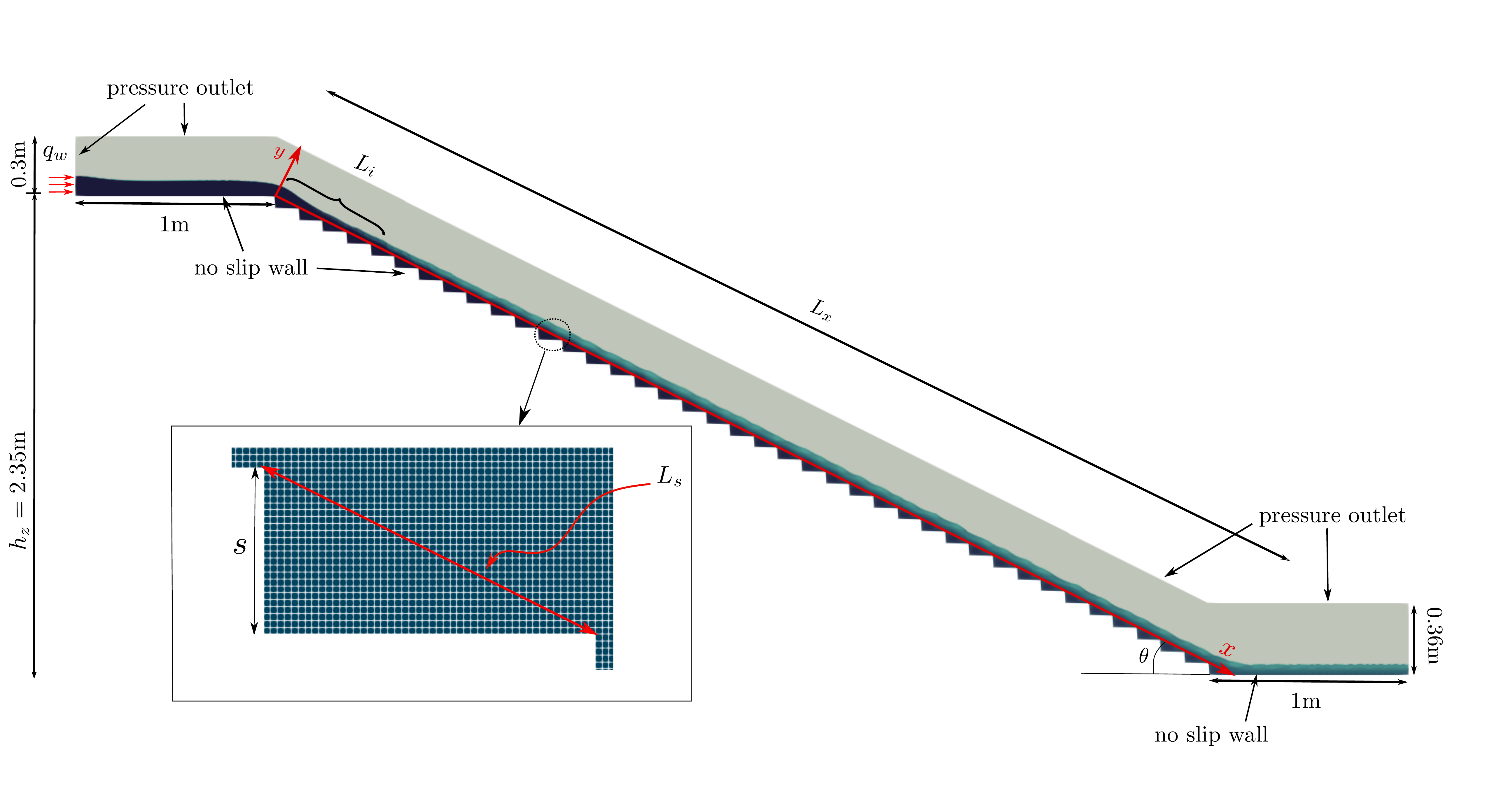}
        \caption{Sketch of the geometry of the simulation case, identifying the  geometric parameters, and also the employed boundary conditions.}
        \label{fig:spillwaySketch}
\end{figure}

This section presents the setup of the stepped spillway simulations used to evaluate the performance of the entrainment modelling.
In order to have a reference with respect to which the accuracy of simulation results can be analysed, the flow and spillway parameters are selected to match those in the experiments of~\citet{Bung2011}.
These were performed on four different spillways combining two selections for the angle ($\theta=$18.4, 26.6\degree) with two for the step height ($s=0.03,0.06$ m).
For each spillway, measurements were made for three flow discharge values ($q=0.07,0.09,0.11$ m$^2$s$^{-1}$).
The parameters $\theta$, $s$, and $q$ can be used to construct the step Froude number, which can be considered the main controlling parameter of the flow~\cite{matos2001,chanson1993stepped},
\begin{equation} \label{eq:froude}
  \textsf{F\textsubscript{s}} = \frac{q}{\sqrt{g\text{ sin}\theta K^3}} .
\end{equation}
Here $K=s\cos\theta$ is the step induced macro-roughness and $g$ is the acceleration due to gravity.
The experiments of Bung cover twelve different Froude numbers in the range, $2.7 \leq \textsf{F\textsubscript{s}} \leq 13$.
For the simulations four values fairly evenly distributed across this range have been selected: $2.7$, $4.6$, $8.3$, and $13.0$.
The values of $\theta$, $s$, and $q$ in the four simulation cases are provided in Table~\ref{tab:fscases}.
This table also provides the values of some auxiliary geometrical parameters, the definition of which can be found in Figure~\ref{fig:spillwaySketch}.
The figure also shows the origin and orientation of the employed Cartesian coordinate system.



\begin{table}[htp]
  \caption{Setup for the different simulation cases. The number of cells are given in $10^3$.}
  \label{tab:fscases}
  \begin{tabular}{lllllllllll}
    \textsf{F\textsubscript{s}}(-) & $\theta$(\degree) & $s$(m) &
    $q$(m$^2$s$^{-1}$)  & $L_x$(m)  & $L_s$(m) & $h_{win}$(m) &
    $n_{cells,G1}$ & $n_{cells,G2}$ & $n_{cells,G3}$ & $n_{cells,G4}$
    \\
    \hline
    2.7  & 26.6 & 0.06 & 0.07 & 5.23 & 0.134 & 0.10 & 89 & 354 & 1 414 & 5
    647 \\
    4.6  & 18.4 & 0.06 & 0.11 & 7.41 & 0.190  & 0.13 & 122  & 482 & 1 917 & 7 646 \\
    8.3  & 18.4 & 0.03 & 0.07 & 7.41 & 0.095 & 0.10 & 119  & 466 & 1 840 & 7 317 \\
    13.0 & 18.4 & 0.03 & 0.11 & 7.41 & 0.095 & 0.13 & 119  & 466 & 1 840 & 7 317 \\
    \hline
  \end{tabular}
\end{table}


All the simulations are performed on 2D domains.
This is chiefly motivated by the fact that the investigated modelling methodology is low-fidelity and most suitable for quick evaluations of the integral characteristics of the flow.
Any 3D effects due to sidewalls are expected to be negligible with respect to the overall accuracy of the flow predictions.
Additionally, using 2D domains it seems to be ensured that, even on dense grids, no air
entrainment is resolved by the VoF. This is shown by the \ttt{interFoam}
computations on grid G4. By contrast, in a 3D setting, it is more likely that some interface perturbations eventually start getting captured.
Investigating the performance of the entrainment modelling in such a scenario is of interest, but lies out of scope of the current work.

The same boundary conditions were used for all cases, with the exception of the discharge $q$ prescribed at the water inlet.
The height of the inlet was adjusted to ensure sub-critical inflow conditions.
A zero gradient condition was used for the pressure, while Dirichlet conditions
were applied to $k$, $\eps$, and $\omega$. The values were set 
assuming 5\% turbulent intensity and 10\% of the critical height as the
turbulent length scale.

For the outlet, a zero gradient condition was prescribed for velocity, pressure,
$\alpha_l$ and $\alpha_g$.
For $k$, $\eps$, and $\omega$ the OpenFOAM \ttt{inletOutlet} boundary condition was applied.
It acts as a zero gradient condition in case of outflow, but for backflow a homogeneous Dirichlet condition is applied instead.

No slip conditions were used at the walls, with a zero gradient condition set for $\alpha_l$ and $\alpha_g$.
The turbulent quantities were estimated by regular wall laws, in OpenFOAM named as
\ttt{kqRWallFunction}, \ttt{epsilonWallFunction}, \ttt{omegaWallFunction} and
\ttt{nutkWallFunction}, for $k$, $\eps$, $\omega$, and $\nu_t$
respectively. 

For the top boundary the total pressure was fixed, and a \ttt{pressureInletOutletVelocity} condition was applied for the velocity.
Similar to \ttt{inletOutlet}, this imposes a zero gradient for outflow, whereas for backflow, it assigns a velocity based on the flux in the patch normal direction.
The \ttt{inletOutlet} boundary condition was used for $\alpha_l$, $\alpha_g$, $k$, $\eps$, and $\omega$.

The material properties of the fluids were set to correspond to air and water.
The values are provided in Table \ref{tab:simpara}.

\begin{table}[htp]
  \centering
  \begin{minipage}{.5\textwidth}
  \caption{Simulation parameters.}
  \label{tab:simpara}
  \begin{tabular}{ll}
    \textbf{Property} & \textbf{Value}  \\
    \hline
    Liquid density, $\rho_1$             & 1000 kg/$\mathrm{m^3}$   \\
    Gas density, $\rho_2$                & 1    kg/$\mathrm{m^3}$   \\
    Liquid kinematic viscosity, $\nu_1$  & $1 \cdot 10^{-6}$ $\mathrm{m^2}/\mathrm{s}$    \\
    Gas kinematic viscosity, $\nu_2$     & $1.48 \cdot 10^{-5}$ $\mathrm{m^2}/\mathrm{s}$ \\
    Surface tension coefficient, $\sigma$       & 0.07        \\
    \hline
  \end{tabular}
  \end{minipage}
\end{table}

The computational grids were constructed using Pointwise\textsuperscript{\textregistered}, and consist of square cells with the exception of a small strip close the top boundary, where unstructured meshing was necessary to account for the slope of the geometry.
Four grids with increasing cell density, denoted G1, G2, G3, and G4, were constructed for each of the four spillways.
In each consecutive grid the edge length of the square cells is halved.
On the coarsest grid G1, the edge length is 5 mm, which corresponds to what was used in the simulations by~\citet{Lopes2017}.
This can be related to the the critical height of the spillway flow, defined as
$h_c = \big( q^2/g\big)^{1/3}$.
Depending on the flow case, on the G1 grid, $h_c$ is discretised by either 15 or 21 cells.
The numbers for the G4 grid are, respectively, 126 and 171.
The densities are not adjusted to remain equal with respect to $h_c$ across all flow conditions, because experiments showed that the relevant parameter for entrainment modelling is the resolution of the interface.
The number of cells in each mesh is given in Table~\ref{tab:fscases}.

In conclusion, additional characteristic scales of spillway flow are defined.
These will be used for non-dimensionalising the results.
At a given $x$, the height $h_{90}$ is defined as the $y$-coordinate of the point where $\alpha_{air}=0.9$.
The velocity $u_{90}$ is defined as the $x$-component of the mean velocity vector at $y=h_{90}$.
Similar scales can be defined with respect to other $\alpha_{air}$ values, e.g.~$h_{50}$.

\section{Air entrainment modelling}
\label{sec:airentmod}

This section presents the air entrainment model developed by~\citet{Lopes2017}.
One can split the model into three components: an estimator for the flux of
entrained air, a transport equation for the volume fraction of entrained air, and a coupling procedure between the model and the VoF framework.
Sections~\ref{subsec:fluxest}, \ref{subsec:alphag}, and \ref{subsec:coupling} each focus on one of these components.
Additionally, for the stepped spillway, estimating the location of the inception point is necessary and this constitutes an additional component of the model, which is treated in Section~\ref{subsec:inceptestaif}.

\subsection{Estimating the flux of entrained air}
\label{subsec:fluxest}

A key component of the model is the estimation of the quantity of entrained air carried passed some imaginary surface located below the interface.
The estimate was proposed by~\citet{ma2011}:

\begin{equation} \label{eq:source_ma}
    q = a \cdot \text{Pos}\left( \nabla (u \cdot n) \cdot n  \right),
\end{equation}
where $a$ is a length scale associated with the roughness of the interface due to turbulence,
\[
\text{Pos}(x) =\left\{
                \begin{array}{ll}
                  x , \quad x \geq 0 \\
                  0 , \quad x < 0,
                \end{array}
                \right.
\]
and $n$ is the interface normal defined as
\begin{equation} \label{eq:n_interface}
  n=\nabla {\alpha_l} / \left( \vert \nabla {\alpha_l}
         \vert + \varepsilon \right).
\end{equation}
Here $\varepsilon$ is a small number added for numerical stability.

It is assumed that entrainment is confined to a layer of thickness $\phi_{ent}$ close to the surface.
Therefore, in order to obtain a volumetric air entrainment rate, $q$ can be divided by $\phi_{ent}$.
Note, however, that \eqref{eq:source_ma} is by definition not restricted to being non-zero only in the vicinity of the interface.
Theoretically, entrainment can be incorrectly predicted in regions where it should not take place.
For this reason, in~\cite{Lopes2017}, $q$ is additionally multiplied by some function $\delta_{fs}$, which is non-zero only close to the interface.
The final form of the volumetric air entrainment rate estimate is

\begin{equation}  \label{eq:sg}
  S_{g} = \frac{a}{\phi_{ent}}
  \text{Pos}\left( \nabla (u \cdot n) \cdot n  \right)  \delta_{fs}.
\end{equation}

It remains to define how $a$, $\phi_{ent}$, and $\delta_{fs}$ are computed.
The common approach for $a$ is to equate it to the turbulent length scale as predicted by the RANS model.
The value of $\phi_{ent}$ should be related to some characteristic length scale
of the problem. 

Within the VoF framework, the $\alpha_l$-field stands out as the natural choice as a basis for the development of an interface indicator function such as $\delta_{fs}$.
Typically, the interface is defined as the isosurface $\alpha_l = 0.5$, however this is only accurate when the interface is sharp.
In the presence of air entrainment, a more robust metric is the magnitude of the gradient of $\alpha$, which can be expected to reach its maximum close to the boundary between the continuous air region and the air-water mixture.
\citet{hansch2012multi} used the gradient of $\alpha$ and a function based on tanh to find the interface as part of their air entrainment model.
This function was adopted by~\citet{Lopes2017} and reads as
\begin{equation}
\delta_{fs} = \frac{1}{2} \text{tanh} \big[ \beta \Delta x
  (|\nabla\alpha_l|-|\nabla\alpha_l|_{cr}) \big] + 0.5.
\label{eq:aifdeltafs}
\end{equation}
Here $|\nabla\alpha_l|_{cr}$ is a constant representing the critical value of the gradient that is expected to be reached in the interface cells.
Its estimate can be computed based on the size of the grid cell, $\Delta x$: $|\nabla\alpha_l|_{cr}={1}/{(4 \Delta x)}$.
The parameter $\beta$ can be used to control the extent of the interface region with respect to the chosen $|\nabla\alpha_l|_{cr}$, and thus provides an opportunity to broaden or restrict the number of cells in which the source term is active.

\subsection{The $\alpha_g$-equation}
\label{subsec:alphag}
The source term~\eqref{eq:sg} is introduced into an additional equation for the modelled volume fraction of entrained air, $\alpha_g$:

\begin{equation}
  \frac{\partial \alpha_g}{\partial t}+
  \nabla \cdot \left(u_{g} \alpha_g \right)
  +   \nabla \cdot \left( \nu_t \nabla {\alpha_g} \right)
= S_g.
  \label{eq:alphag}
\end{equation}
Here $\nu_t$ denotes the turbulent viscosity. The velocity of the entrained air, $u_{g}$, is either set to be equal to $\overline{u}$ or alternatively modified according to~\cite{Clift1978}:
\begin{equation}
  {u}_{g} = \overline{u} + {u}_{r},
  \label{eq:ug}
\end{equation}
where the correction velocity ${u}_{r}$ is calculated based on a bubble
radius according to
\begin{equation}
      {u}_{r}=\left\{
                \begin{array}{ll}
                  -4474r_b^{1.357}g, \quad \text{if} \quad 0 < r_b \leq
                  7 \times 10^{-4} \text{m} \\
                  \\
                  -0.23{g}, \quad \text{if} \quad 7 \times 10^{-4} < r_b \leq
                  5.1 \times 10^{-3} \text{m} \\
                  \\
                  -4.202r_b^{0.547}g \text{ if } r_b >
                  5.1 \times 10^{-3} \text{m}.
                \end{array}
                \right.
      \label{eq:ur}
\end{equation}
The inclusion of the diffusion term in (\ref{eq:alphag}) is considered optional.

Additionally,~\citet{Lopes2017} argue that to properly account for the break-up of bubbles at the free surface, $\alpha_g$ should be set to zero when $\alpha_{air}$ exceeds a certain threshold value, referred to as the BBA. The suggested value to use is $0.1$.

The exact physical meaning of $\alpha_g$ and its relation to $\alpha_{l}$ are somewhat elusive.
In~\citep{Lopes2017}, the authors discuss the possibility of using the entrainment model without backward coupling to the VoF solver.
In this case, the situation is clear: $\alpha_g$ shows the modelled distribution of the volume fraction of entrained air, which cannot be captured by the VoF.
However, when the coupling is two-way (particulars presented below), the idea is that the entrained air should be captured in the $\alpha_{l}$ field, and $\alpha_g$ is essentially reduced to a buffer-field used to propagate the effect of $S_g$ onto $\alpha_{l}$.

\subsection{Coupling to the VoF solver} \label{subsec:coupling}
Here, we are interested in applying the model in a two-way coupling regime, meaning that the model's predictions should be propagated into the distribution of $\alpha_{l}$.
The premise is that the VoF simulation by itself does not resolve any entrainment, and therefore all of it is accounted for by a subgrid model based on the $\alpha_g$ equation~\eqref{eq:alphag}.
The overall idea is that $\alpha_l$ should be reduced in regions where $\alpha_g$ is large, and in a manner that does not disrupt mass conservation.

Here this is done through a modification of the artificial compression term introduced into the $\alpha_l$-equation~(\ref{eq:alpha_mules}):

\begin{equation}  \label{eq:compressionTerm}
  \nabla \cdot \left(u^c(1 -\alpha_l)\alpha_l\right).
\end{equation}
The term $(1 -\alpha_l)\alpha_l = \alpha_{air}\alpha_{l}$ is originally meant to serve as an indicator for cells constituting the interface, in which the compression is to be applied.
The key observation is that multiplying $u^r_j$ by some negative number instead would lead to interface expansion and thus a region occupied by a mixture.
The goal is then to correlate $\alpha_g$ with the change in sign in the term in front of $u^r_j$.
The most obvious way to do that is exchange $\alpha_{air}\alpha_l$ for $(\alpha_{air} - \alpha_g)\alpha_l$.
Note that since~\eqref{eq:compressionTerm} is a transport term, mass conservation is guaranteed.
The modified $\alpha_l$-equation then reads
\begin{equation}
\label{eq:alpha_mules_aif}
    \pdiff{\alpha_l}{t} + \nabla \cdot (\overline u \alpha_l ) +
    \nabla \cdot \left(u^c (\alpha_{air} - \alpha_g)\alpha_l\right) = 0.
\end{equation}
Under the definition above, the model is active only when $\alpha_g > \alpha_{air}$, which is reasonable.
It is also worth mentioning that otherwise~\eqref{eq:compressionTerm} recovers its original compressive function.
This occurs even in the regions occupied by a mixture, which can be called into question.
As part of the work on improving the model, some experiments have been conducted in which positive values of $\alpha_{air} - \alpha_g$ where cut to 0, however the exhibited results were inaccurate, and introducing such a discontinuity is probably best avoided.




\subsection{Inception point estimation}
\label{subsec:inceptestaif}

As discussed in the introduction, surface aeration initiates when the turbulence perturbations exceed the stabilizing forces of surface tension and buoyancy at the free surface.
In the model of \citet{Lopes2017} no attempt is made to explicitly compute the force balance.
Instead, two model parameters, $k_c$ and $u_c$ are introduced, where the subscript $c$ stands for critical.
The inception is considered to occur when
\begin{align}
 \label{eq:kc}
  k > k_c \; \text{and} \;  u \cdot n > u_c \; \text{and} \; u \cdot g > u_c.
\end{align}

Appropriate values for $k_c$ and $u_c$ are extremely difficult to predict a priori, since the selection clearly depends not only on the flow conditions (see Section~\ref{sec:aiffscomp}), but also on the turbulence model and its prediction of $k$.
Careful calibration with respect to the selected model is therefore necessary.
In~\citep{Lopes2017}, the authors nevertheless suggest $u_c = 0.8$ m/s and $k_c = 0.2$ $\text{m}^2/\text{s}^2$, referring to previous experimental results.
Unfortunately, how these values relate to the characteristic length and velocity scales of the flow is not discussed.

\section{Stepped spillway simulations with the original model}
\label{sec:aifresults}
This section demonstrates results from simulations performed using the model of
\citet{Lopes2017} described in the previous section. 
In the original source the model was reported to reproduce experimental data on
a stepped spillway with \textsf{F\textsubscript{s}}=2.7. However, the
simulations were performed on relatively coarse grids, with the grid sensitivity
study performed using only the baseline VoF solver. Furthermore, initial testing within this work
indicated that its effect was reduced upon grid refinement, which motivates the
analyses herein.

The implementation of the corresponding solver, called \ttt{airInterFoam} was kindly provided by P.~Lopes via personal communication.
Below, we abbreviate \ttt{airInterFoam} to AIF.
In the following sections, the solver is evaluated in terms of sensitivity to flow conditions (Section~\ref{sec:aiffscomp}), grid resolution (Section~\ref{sec:gridsens}), and several parameters of the entrainment model (Section~\ref{sec:aifur}).

\subsection{Sensitivity to \textsf{F\textsubscript{s}}}
\label{sec:aiffscomp}

Here, results from four AIF simulations varying in the prescribed Froude number of the spillway flow are presented.
To highlight the effect of the entrainment model, data obtained with the baseline VoF solver \texttt{interFoam} (abbreviated IF henceforth) are also provided.
The employed numerical parameters are based on~\citep{Lopes2017}, where good results for the \textsf{F\textsubscript{s}}=2.7 case (simulated in 3D) are presented.
In particular, the G1 grid is employed, $k$-$\omega$ SST is used for turbulence
modelling, the diffusion term is omitted in the $\alpha_g$ equation, and $k_c =
0.2$ m$^2$/s$^{2}$, $u_c=0$, $u_g=u_l$, $\text{BBA} = 0.1$, and $\phi_{ent}=0.05h_c$.

\begin{figure}[htp]
   \vspace{-8pt}
   \centering
      \subcaptionbox{\textsf{F\textsubscript{s}}=2.7  \label{subfig:fs27_alpha}}
        {\includegraphics[scale=1,trim={0.0 10 0 0},clip]{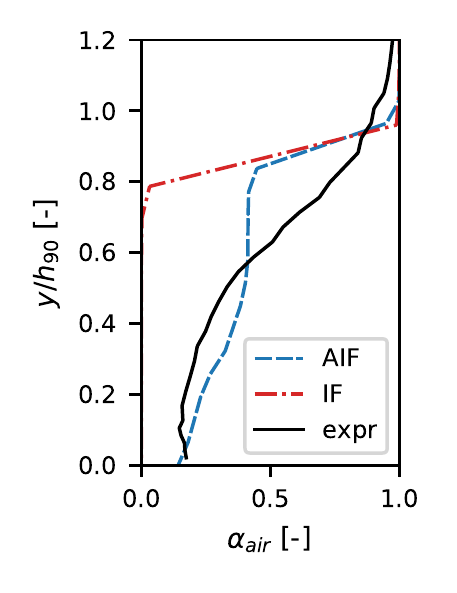}}
      \subcaptionbox{\textsf{F\textsubscript{s}}=4.6\label{subfig:fs46_alpha}}
                    {\includegraphics[scale=1,trim={0.8cm 10 0 0},clip]{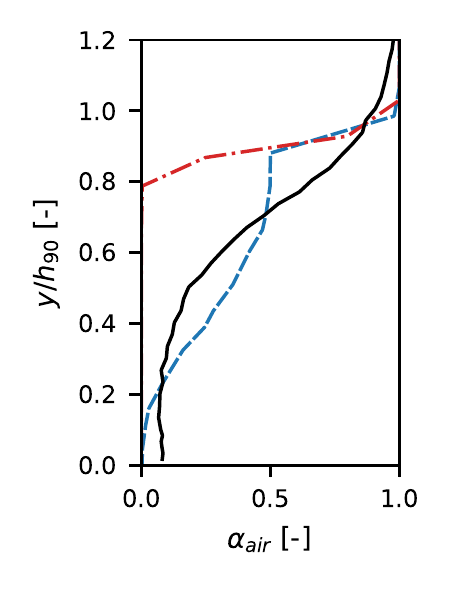}}
      \subcaptionbox{\textsf{F\textsubscript{s}}=8.28\label{subfig:fs828_alpha}}
        {\includegraphics[scale=1,trim={0.8cm 10 0
              0},clip]{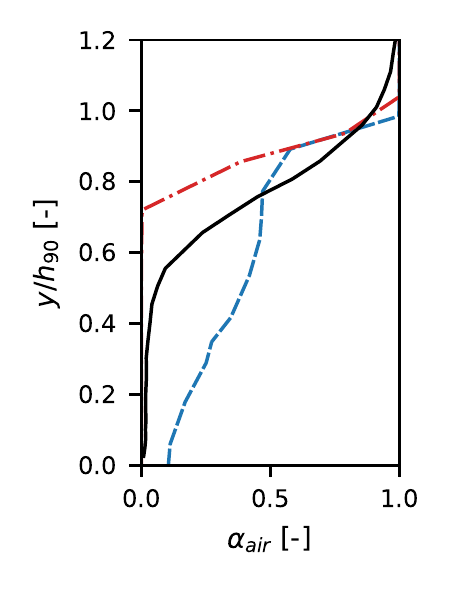}}
      \subcaptionbox{\textsf{F\textsubscript{s}}=13\label{subfig:fs13_alpha}}
        {\includegraphics[scale=1,trim={0.8cm 10 0
              0},clip]{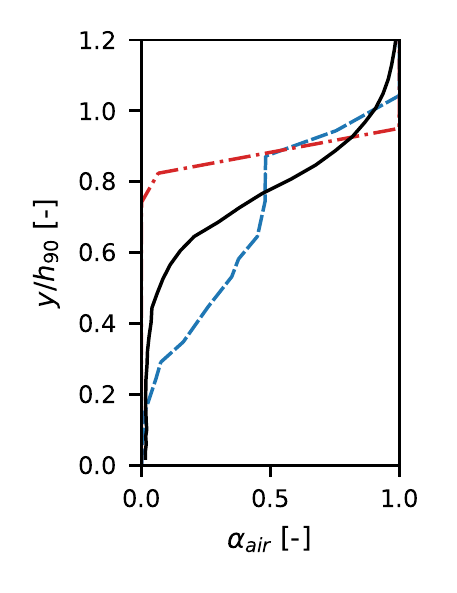}}
        \\
        \caption{Vertical void fraction profiles for uniform flow conditions. Spillway flows with different Froude numbers at the coarsest grid G1. AIF simualtions compared to IF simulations and experimental results by~\citet{Bung2011}.}
        \label{fig:fsComp_alpha_dx005}
\end{figure}

Figure~\ref{fig:fsComp_alpha_dx005} shows the obtained values of $\alpha_{air}$ in the uniform flow region.
Good accuracy is achieved for \textsf{F\textsubscript{s}}=2.7 and $4.6$, but for the two higher $\textsf{F\textsubscript{s}}$ the model fails to predict the reduced penetration of air into the corners of the steps.
As a result, in terms of magnitude, the errors in the IF and AIF simulations are similar, although in the case of IF the diffusion of the interface is a purely numerical effect.
It is also interesting to note that for \textsf{F\textsubscript{s}}=2.7 the accuracy is on par with the 3D simulations using similar model settings conducted in~\citep{Lopes2017}.

\begin{figure}[htp]
  \vspace{-20pt}
   \centering
    \begin{subfigure}[b]{\textwidth}
        \includegraphics[scale=1,trim={0 20 0 0},clip]{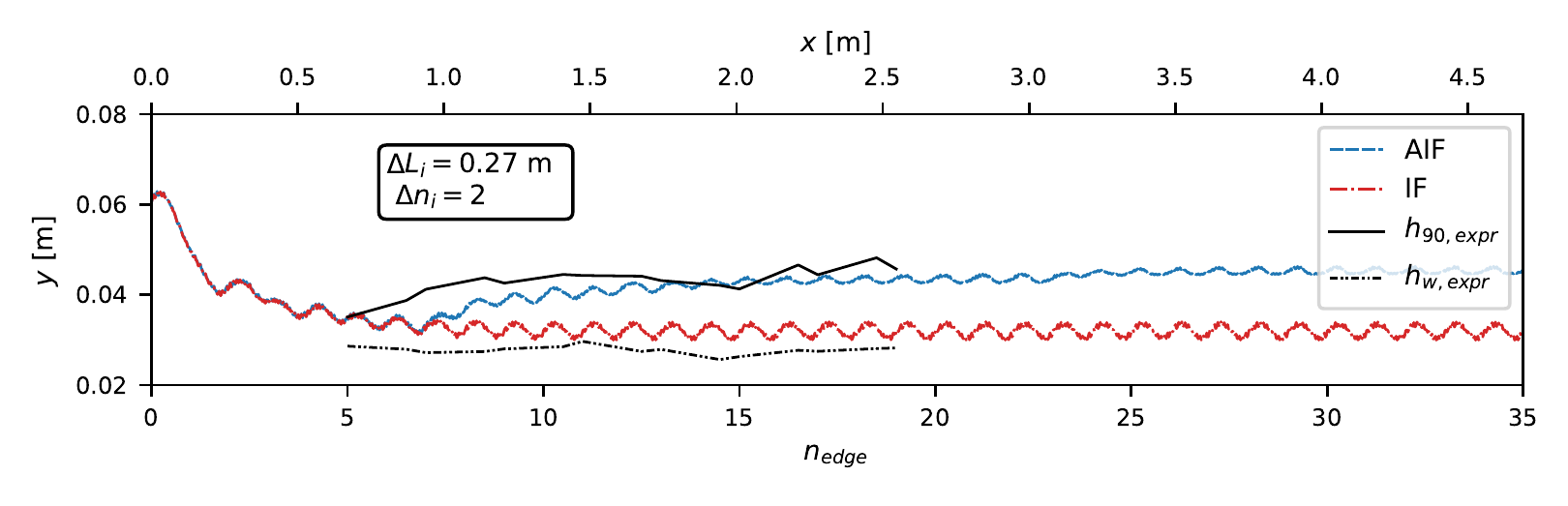}
        \caption{\textsf{F\textsubscript{s}}=2.7}
        \label{subfig:h90_fs27_dx005}
    \end{subfigure}
    \par\medskip
    \begin{subfigure}[b]{\textwidth}
        \includegraphics[scale=1,trim={0 20 0 20},clip]{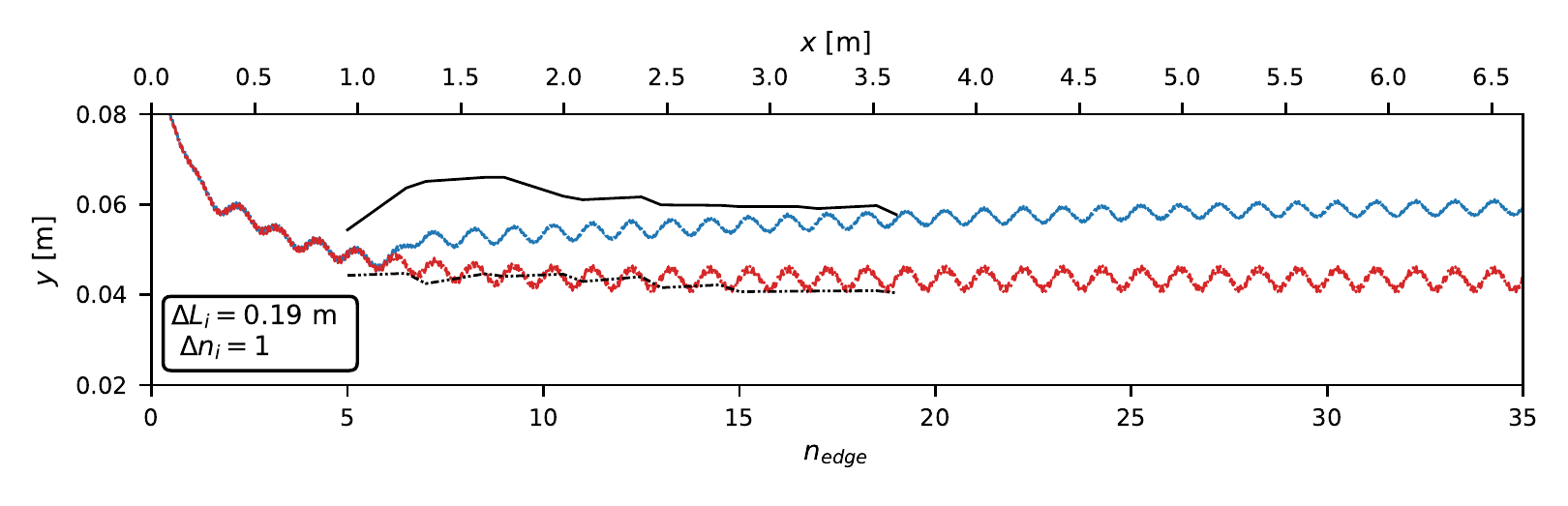}
        \caption{\textsf{F\textsubscript{s}}=4.6}
        \label{subfig:h90_fs46_dx005}
    \end{subfigure}
    \par\medskip
    \begin{subfigure}[b]{\textwidth}
        \includegraphics[scale=1,trim={0 20 0 20},clip]{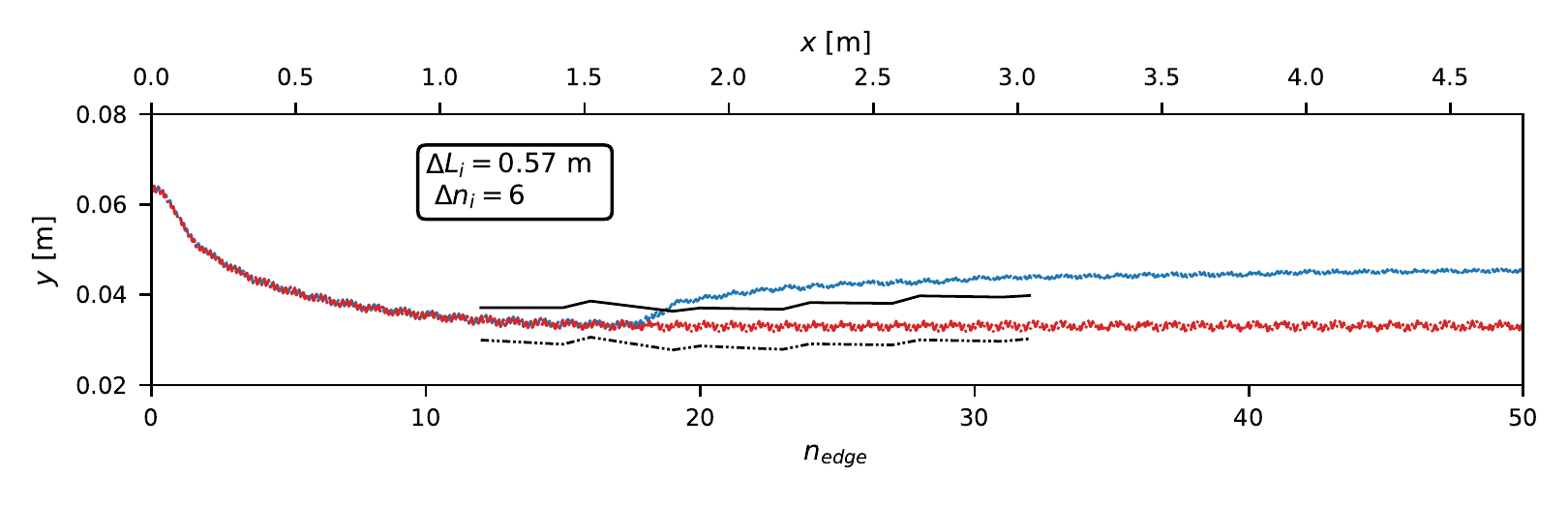}
        \caption{\textsf{F\textsubscript{s}}=8.28}
        \label{subfig:h90_fs82_dx005}
    \end{subfigure}
    \par\medskip
        \begin{subfigure}[b]{\textwidth}
        \includegraphics[scale=1,trim={0 10 0 20},clip]{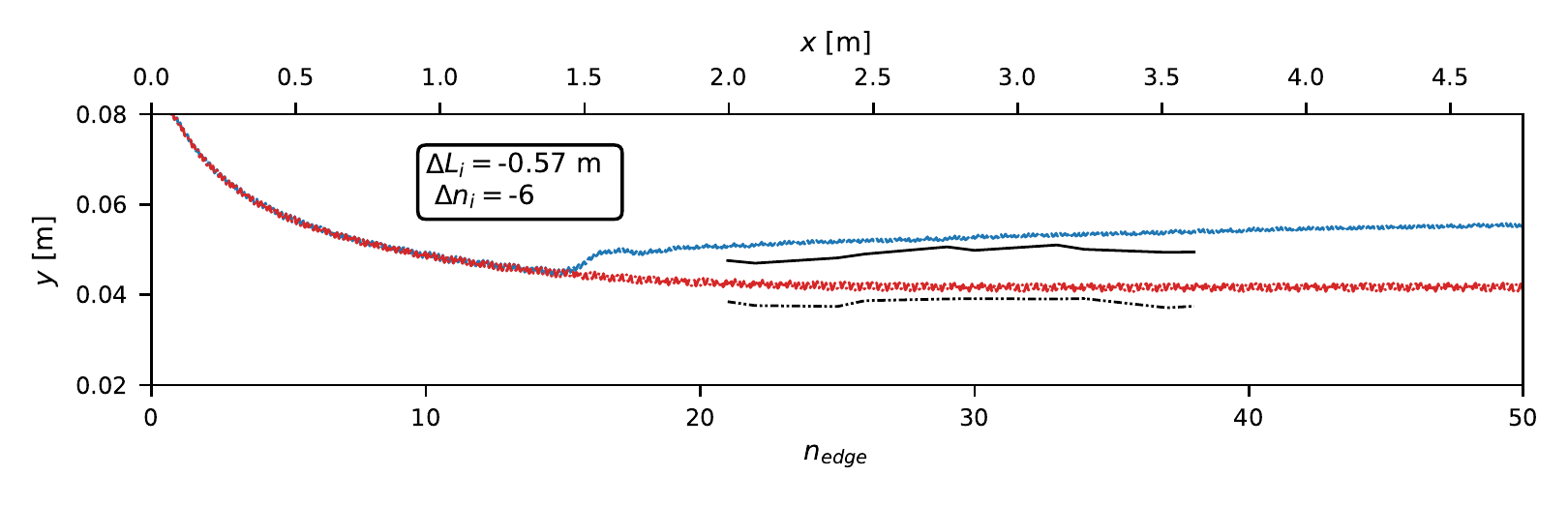}
        \caption{\textsf{F\textsubscript{s}}=13}
        \label{subfig:h90_fs13_dx005}
    \end{subfigure}
    \caption{Surface elevation plots, $h_{90}$. Spillway flows with different Froude
      numbers simulated by AIF on the coarsest grid G1, compared
      to IF simulations and physical model results by \citet{Bung2011}. The
      difference from the experimentally measured inception
      point is annotated in meters ($\Delta L_{i}$) and in steps
      ($\Delta n_{i}$).}\label{fig:h90_fscomp_dx005}
\end{figure}

The evolution of the surface elevation, measured as $h_{90}$, is shown in Figure~\ref{fig:h90_fscomp_dx005}.
The elevation's value in uniform conditions is well-predicted for all Froude numbers.
However, the location of the inception points are not captured as consistently.
The difference in the obtained values with respect to the experimental data of Bung
is provided in each plot of the figure: $\Delta n_{i}$ stands for the difference in terms the step number, and $\Delta L_i$ in terms of $x$.
It should be noted that when comparing across different $F_s$, using $\Delta
L_i$ is more appropriate, since for the two higher Froude numbers, the length of
the step is halved relatively to the lower Froude number cases. 
The obtained incetion point locations for \textsf{F\textsubscript{s}}=2.7 and \textsf{F\textsubscript{s}}=4.6 are reasonably accurate, but, unfortunately, at higher \textsf{F\textsubscript{s}} the disagreement with the experiment becomes larger.
Furthermore, the predicted inception point for \textsf{F\textsubscript{s}}=8.3 is further downstream as compared to that for \textsf{F\textsubscript{s}}=13, whereas the experimental data exhibits the opposite trend.

Figure~\ref{fig:h90_fscomp_dx005} additionally shows the experimental values of $h_w$, which is the equivalent clear water depth, i.e.~the surface elevation that should be predicted by IF.
In the obtained results, IF somewhat over-predicts $h_w$, the reason being the coarseness of the grid.


\begin{figure}[htp]
   \centering

       \subcaptionbox{\textsf{F\textsubscript{s}}=2.7\label{fs27_vel}}
         {\includegraphics[scale=1,trim={0 10 0 0},clip]{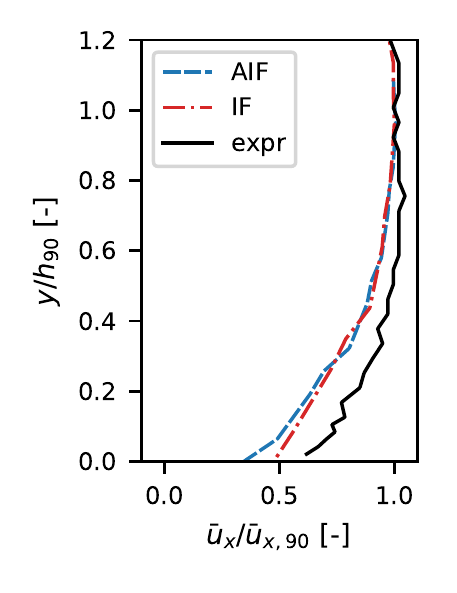}}
       \subcaptionbox{\textsf{F\textsubscript{s}}=4.6\label{fs46_vel}}
                     {\includegraphics[scale=1,trim={0.8cm 10 0 0},clip]{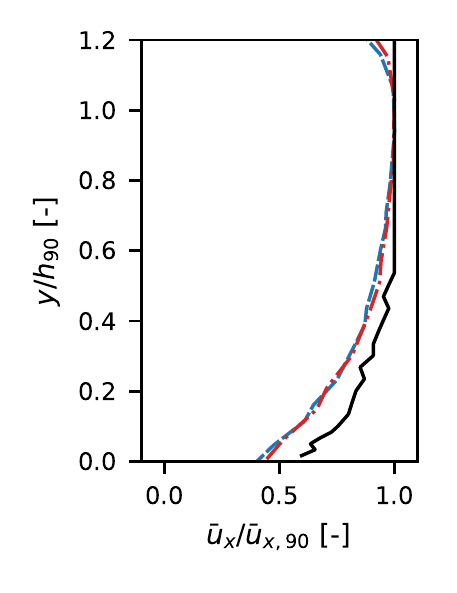}}
       \subcaptionbox{\textsf{F\textsubscript{s}}=8.28\label{fs828_vel}}
         {\includegraphics[scale=1,trim={0.8cm 10 0
               0},clip]{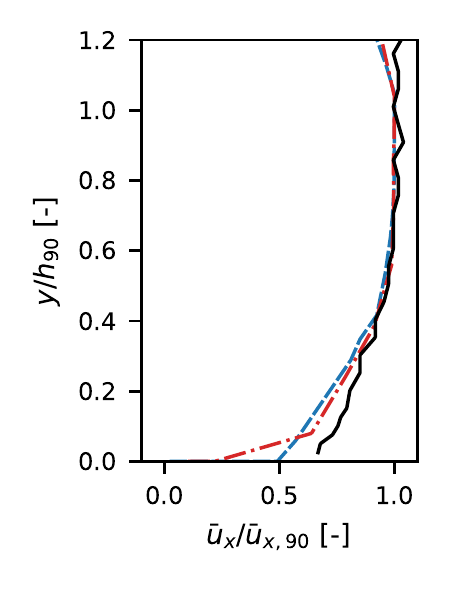}}
         \subcaptionbox{\textsf{F\textsubscript{s}}=13\label{fs13_vel}}
         {\includegraphics[scale=1,trim={0.8cm 10 0
               0},clip]{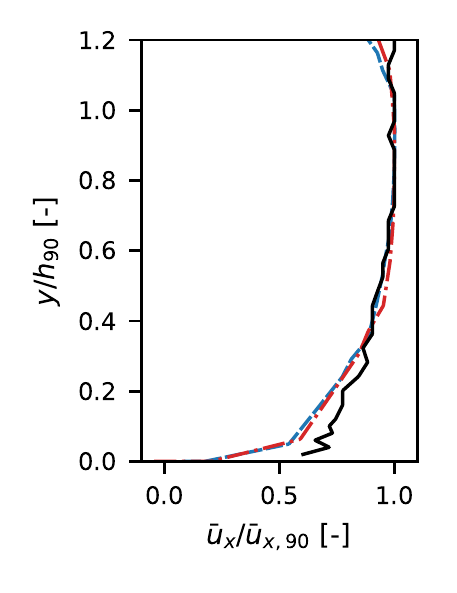}} \\
         \caption{Vertical profiles for the streamwise velocity for uniform flow
           conditions. AIF and IF simulation results compared to experimental
           results by~\citet{Bung2011}.}\label{fsComp_vel} 
 \end{figure}

The predicted profiles of the streamwise velocity are shown in Figure~\ref{fsComp_vel}.
Remarkably, no effect of air entrainment modelling is visible, and accurate profiles can be predicted with IF.
This result was reproduced in all the simulations in this paper, and, for that reason, velocity profiles are not further presented or discussed.

The principle conclusion from the obtained results is that the settings used in~\citep{Lopes2017} for the \textsf{F\textsubscript{s}}=2.7 case fail to provide consistently accurate results as the Froude number becomes larger.
This indicates that some parameter values of the model, for example $k_c$, should be made a function of \textsf{F\textsubscript{s}}.

\subsection{Grid sensitivity}
\label{sec:gridsens}

\begin{figure}[htp]
   \vspace{-7pt}
   \centering
   \subcaptionbox{Surface elevation plot, $h_{90}$-surface.\label{subfig:h90_fs27_aif_grid}}
        {\includegraphics[scale=1,trim={0.4cm 10 0.2cm
              3},clip]{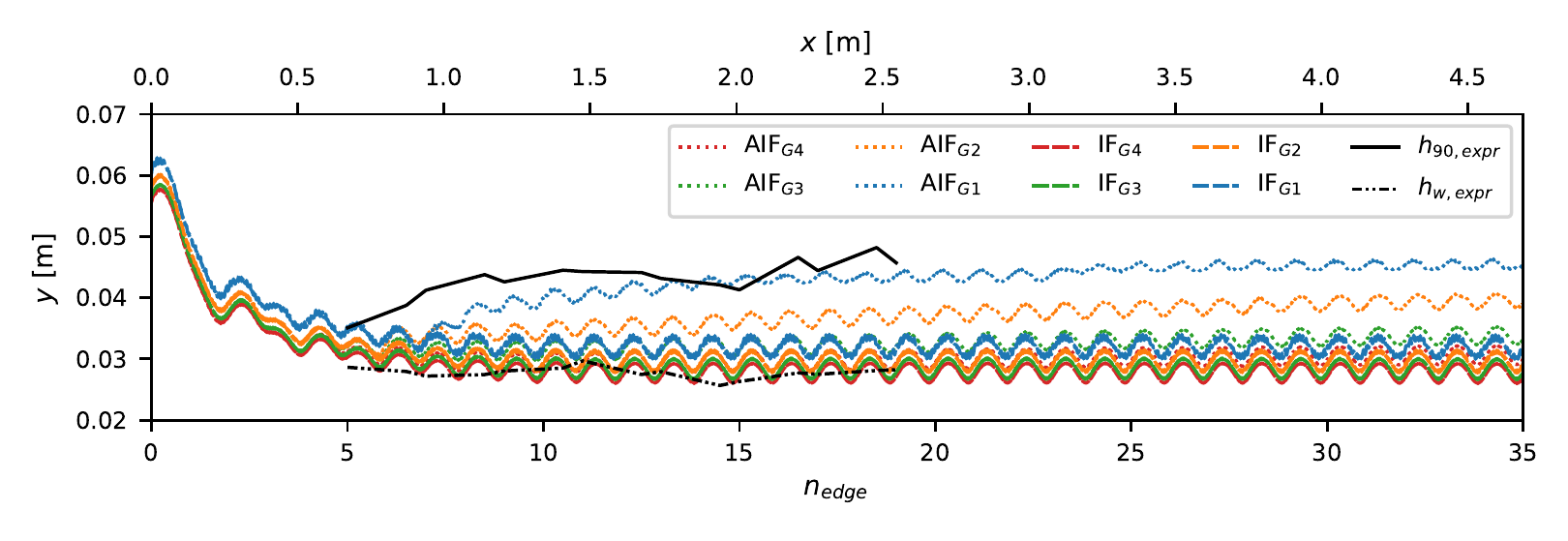}}

      \subcaptionbox{Step 7\label{subfig:gridComp_alpha_M7}}
        {\includegraphics[scale=1,trim={0.1cm 10 0.2cm 3}]{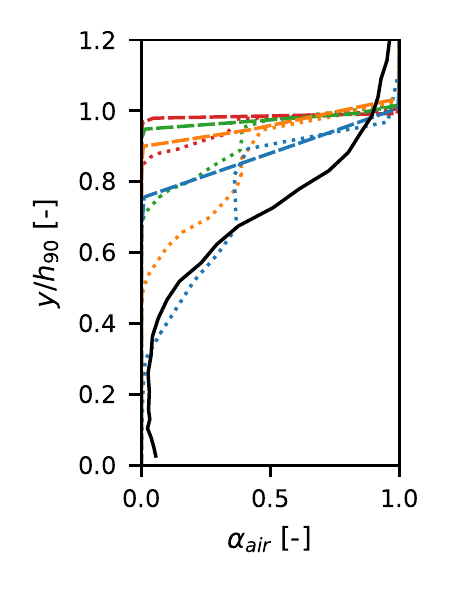}}
      \subcaptionbox{Step 11\label{subfig:gridComp_alpha_M11}}
        {\includegraphics[scale=1,trim={0.7cm 10 0.2cm 3},clip]{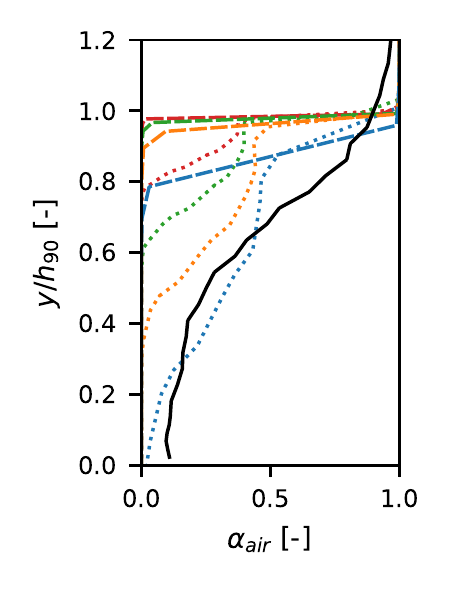}}
      \subcaptionbox{Step 15\label{subfig:gridComp_alpha_M15}}
        {\includegraphics[scale=1,trim={0.7cm 10 0.2cm
              3},clip]{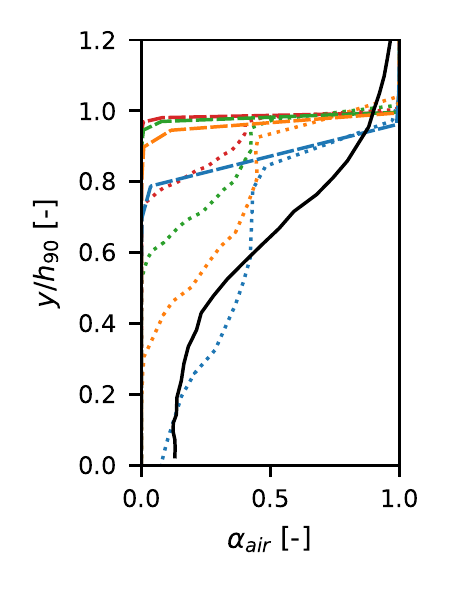}}
      \subcaptionbox{Step 19\label{subfig:gridComp_alpha_M19}}
        {\includegraphics[scale=1,trim={0.7cm 10 0
              3},clip]{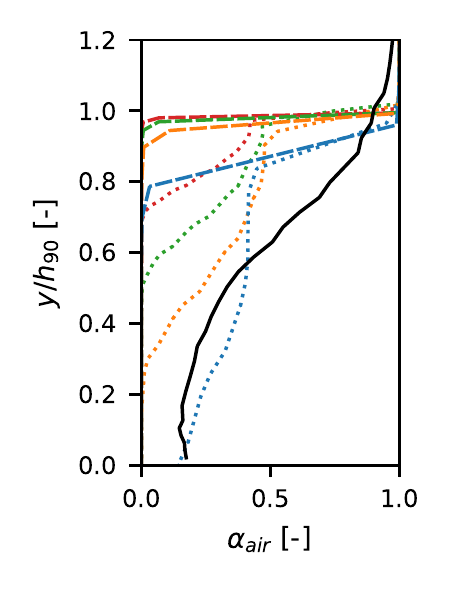}}
        \caption{
          AIF simulations for \textsf{F\textsubscript{s}}=2.7 at
          different grids (G1-G4) compared to 
          physical model results by
          \citet{Bung2011}. Figure~\ref{subfig:h90_fs27_aif_grid} shows the surface
          elevation, Figures
          \ref{subfig:gridComp_alpha_M7}-\ref{subfig:gridComp_alpha_M19}
          void fraction profiles for different steps.
      }\label{fig:aifgridComp}
\end{figure}

To explore the grid sensitivity of AIF, the \textsf{F\textsubscript{s}}=2.7-case was run on all four grids G1-G4.
The resulting $\alpha_{air}$ and $h_{90}$ profiles are shown in Figure~\ref{fig:aifgridComp}.
Clearly, the behaviour obtained on the coarse grid G1 is substantially changed as the grid gets denser.
With increasing resolution, less air is distributed towards the pseudo-bottom, and for the densest grid only a tiny air layer is found close to the surface.
The profiles, both $h_{90}$ and $\alpha_{air}$, approach the corresponding solutions obtained with IF.
Obviously, this is caused by the fact that less numerical diffusion contribute to the
transport of $\alpha_g$ as the grid is refined,
but inspection shows that this is also caused by the shrinkage of the area in
which the source term $S_g$ is non-zero.
This, in turn, is controlled by $\delta_{fs}$, which makes the definition of this
function a contributor to the grid sensitivity.
A more elaborate discussion follows in Section~\ref{sec:modeldevel}.
On the other hand, with respect to $h_w$, the IF solution consistently improves with grid refinement.
On the G4 grid the interface is perfectly sharp and the $h_w$ profile is very well-matched.

\subsection{Sensitivity to $u_r$, bubble breakup criterion, and $\alpha_g$-diffusion}
\label{sec:aifur}

Here we explore the effects of the entrainment model parameters that could arguably be considered non-essential or optional.
First, the impact of the air bubble drift velocity model, as given in eq.~\eqref{eq:ur}, is investigated.
This is followed by an analysis of the bubble breakup criteria, BBA.
Finally, the model is tested in terms of activation of the diffusion term in the $\alpha_g$-equation~\eqref{eq:alphag}.
The rest of the numerical setup is similar to that used in the Froude number sensitivity study, see Section~\ref{sec:aiffscomp}.

\begin{figure}[htp]
   \vspace{-8pt}
   \centering
      \subcaptionbox{\label{subfig:urComp_dx0025_alpha_M19}}
        {\includegraphics[scale=1,trim={0.3cm 10 0.3cm
              3},clip]{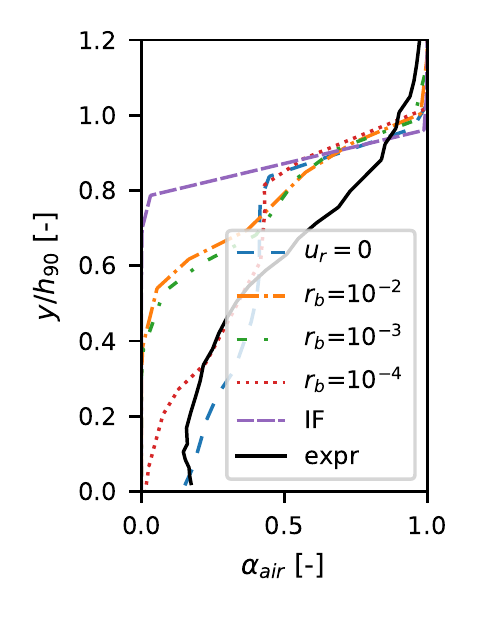}}
      \subcaptionbox{\label{subfig:bbaComp_dx0025_alpha_M19}}
        {\includegraphics[scale=1,trim={0.7cm 10 0.3
              3},clip]{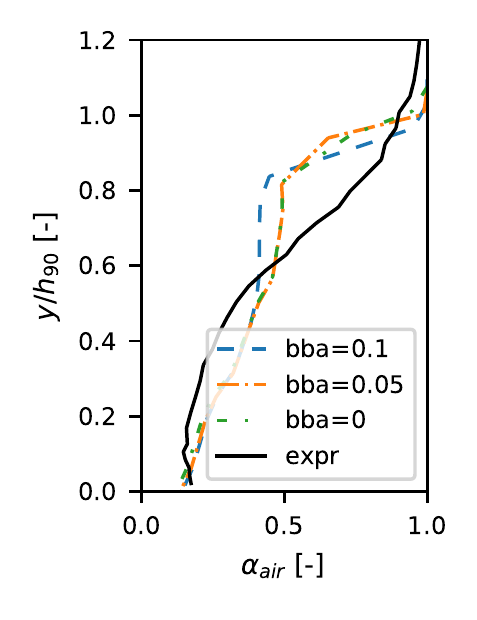}}
        \subcaptionbox{\label{diffaif_dx0025_alpha_M19}}
        {\includegraphics[scale=1,trim={0.7cm 10 0.3cm
              3},clip]{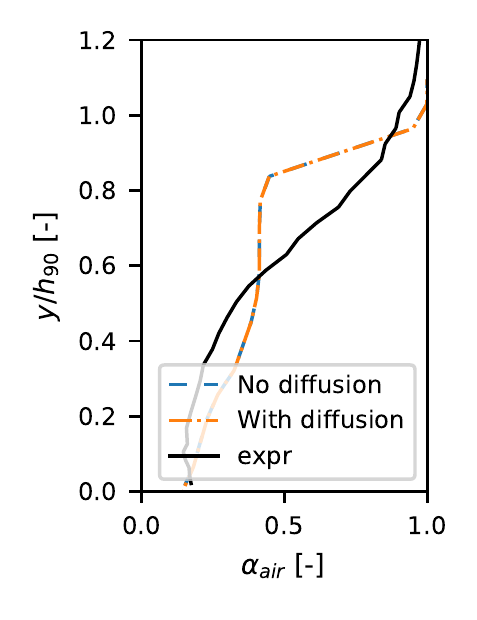}}

        \caption{Sensitivity of $\alpha_{air}$ profiles to the slip velocity model, bubble breakup criterion, and $\alpha_g$-diffusion.}
        \label{fig:urComp_dx005dx0025_alpha}
\end{figure}

Here we restrict the analysis to $\alpha_{air}$ profiles in the uniform flow region.
For the $u_r$ model, three values of the bubble radius $r_b$ are considered, along with setting $u_r$ to 0.
The result is shown in Figure~\ref{subfig:urComp_dx0025_alpha_M19}.
Clearly, including $u_r$ leads to reduced air entrainment, and this effect becomes larger when the input bubble radius is increased.
This is expected, since $u_r$ is, by definition, directed upwards.
Since the intensity of air entrainment is already heavily dependent on the formulation of $\delta_{fs}$, having an additional controller in terms of $u_r$ introduces unnecessary complication.
At least in the case of spillway flow, setting $u_r = 0$ can therefore be recommended.

Three values of the BBA are considered.
The first is 0.1, which is recommended by~\citet{Lopes2017}, and the other two are 0.05 and 0.
Recall that the chosen value refers to the minimal volume fraction of water for which $\alpha_g$ is allowed to be non-zero.
The results from the three simulations are shown in Figure~\ref{subfig:bbaComp_dx0025_alpha_M19}.
As expected, slightly more air is entrained close to the $h_{90}$ when the bba value is reduced.
Note that compared to the experimental data, the air fractions predicted by AIF close to the interface (from $y/h_{90}\approx 0.8$) are too low.
Thus, deactivating the BBA-criterion produces a small improvement in the accuracy meaning that this parameter can be safely removed from the formulation of the model.

Finally, the effect of the diffusion term in~\eqref{eq:alphag} is investigated.
Two simulations, with and without the diffusion term included, have been conducted, see Figure~\ref{diffaif_dx0025_alpha_M19}.
The effect of the diffusion appears to be completely negligible.
Consequently, it is possible to remove it from the model.

In summary, $u_r$, BBA, and $\alpha_g$-diffusion can be excluded from the model, significantly simplifying its formulation.

\subsection{Sensitivity to $\beta$}
\label{sec:aifbeta}

As shown in Section~\ref{sec:gridsens}, at high grid resolutions the effects of
the model becomes negligible, or even deactivated. 
The most important controller of the breadth of the region (in terms of $\nabla \alpha_l$) where entrainment is introduced is the form of $\delta_{fs}$, see Eq.~\eqref{eq:aifdeltafs}.
In particular, the parameter $\beta$ determines the breadth of the tanh function, meaning that with smaller $\beta$ the region of the model activation becomes larger.
In principle, it may thus be possible to maintain an appropriate level of entrainment at high grid resolutions by adjusting $\beta$ accordingly.
However, for this to be possible in practice, the necessary change to $\beta$ should be easy to predict a priori.

Multiple simulations across different Froude numbers and grid resolutions have been conducted in an attempt to determine whether clear guidelines for setting $\beta$ could be established.
Unfortunately, these efforts were fruitless, and the results of using a given $\beta$ value change significantly depending on the flow and parameters of the simulation.

\begin{figure}[htp]
   \vspace{-10pt}
   \centering
   \subcaptionbox{\textsf{F\textsubscript{s}}=2.7\label{subfig:fs27_alpha_betatest}}
                 {\includegraphics[scale=1,trim={0.cm 11 0
                       0},clip]{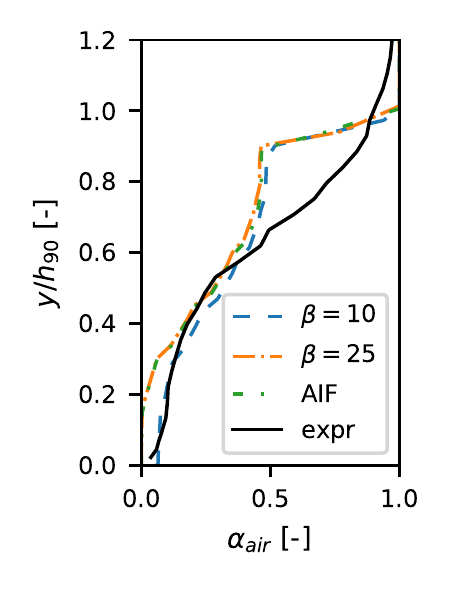}}
   \subcaptionbox{\textsf{F\textsubscript{s}}=4.6\label{subfig:fs46_alpha_betatest}}
                 {\includegraphics[scale=1,trim={0.8cm 11 0
                       0},clip]{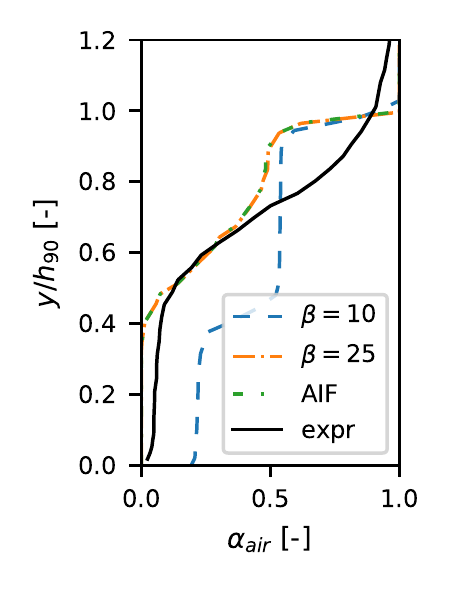}}
   \subcaptionbox{\textsf{F\textsubscript{s}}=8.28\label{subfig:fs82_alpha_betatest}}
                 {\includegraphics[scale=1,trim={0.8cm 11 0
                       0},clip]{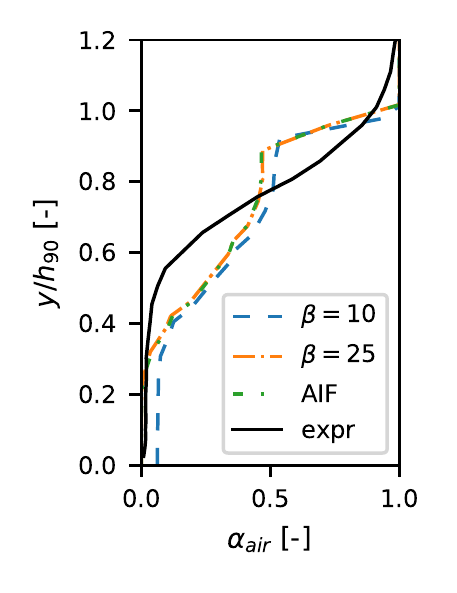}}
   \subcaptionbox{\textsf{F\textsubscript{s}}=13\label{subfig:fs13_alpha_betatest}}
                 {\includegraphics[scale=1,trim={0.8cm 11 0
                       0},clip]{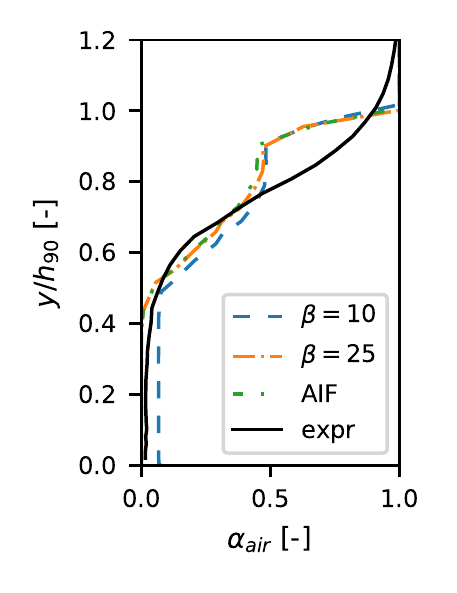}}
        \\
        \vspace{-5pt}
        \caption{Sensitivity on $\beta$ for AIF evaluated on grid
          G2 for different step Froude numbers. Vertical $\alpha_{air}$ profiles are
          shown. The data is extracted from what should be the start of the
          uniform flow region according to the experimental
          data~\citep{Bung2011}. 
          }\label{fig:betacomp_alpha_sE_g2_aif}
\end{figure}

As an illustrative example, Figure~\ref{fig:betacomp_alpha_sE_g2_aif} shows the
$\alpha_{air}$ profiles produced using $\beta = 10$ and $25$ in simulations at
different Froude numbers on the G2 grid,
  and compares it to the corresponding
  profile produced by AIF, where a $\beta$-value of 100 is used as default.
For \textsf{F\textsubscript{s}}=2.7 the sensitivity to $\beta$ is rather small,
but for \textsf{F\textsubscript{s}}=4.6 a significant increase in aeration
occurs.
For larger Froude numbers a lower $\beta$ mainly leads to more air being
present in the corner of the steps. 
While the change in the $\alpha_{air}$ profiles appears to be rather small,
uniform flow conditions are not reached, and its effect continues to grow
further downstream. 
Similar sporadic behaviour was observed with respect to other simulation parameters.
For example, contrary to what is observed when using the G2 grid, using G3
instead leads to the \textsf{F\textsubscript{s}}=2.7 case becoming very
sensitive $\beta$.

\section{Proposed model developments}
\label{sec:modeldevel}
\noindent

In the previous sections, two issues with the entrainment model proposed by~\citet{Lopes2017} have been identified.
Perhaps the most critical one is the successive deactivation of the model upon
grid refinement.
The other one is the difficulty in prescribing the $k_c$ value in order to get a good prediction of the inception point.
In this section, improvements to the model are proposed aiming at alleviating these problems.
First, an alternative formulation for the $\delta_{fs}$ function is introduced in Section~\ref{subsec:fsdetection}.
Afterwards, amplification of the diffusion term in the $\alpha_g$ is argued for in Section~\ref{subsec:extradiff}.
Finally, a different way of predicting the inception point is discussed in Section~\ref{subsec:sgact}.

%

\subsection{Free surface detection}
\label{subsec:fsdetection}

\begin{figure}[htp]
   \vspace{-8pt}
   \centering
      \includegraphics[width=0.58\textwidth]{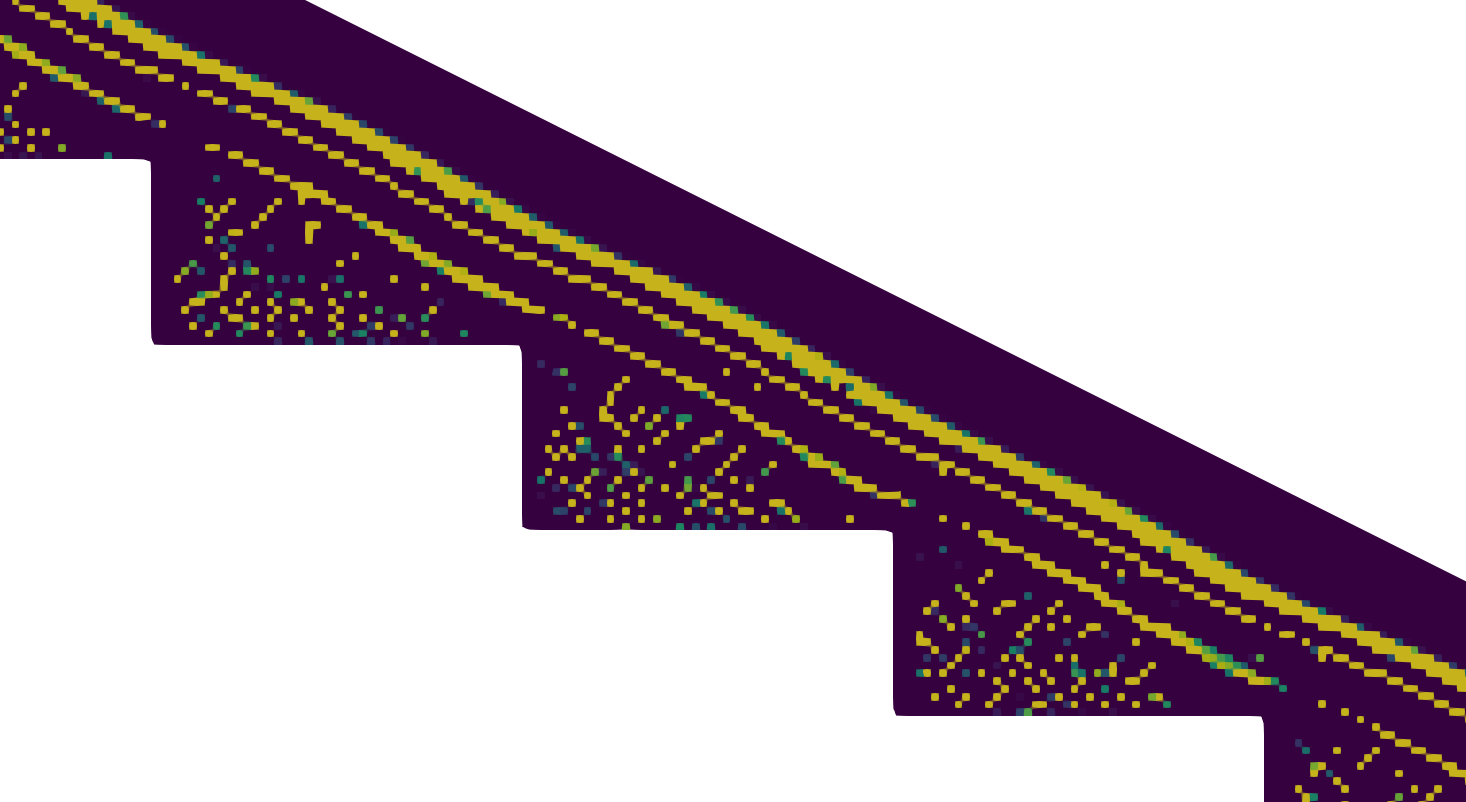}
      \includegraphics[scale=1,trim={0.3cm 10 0.2cm 0}, clip]{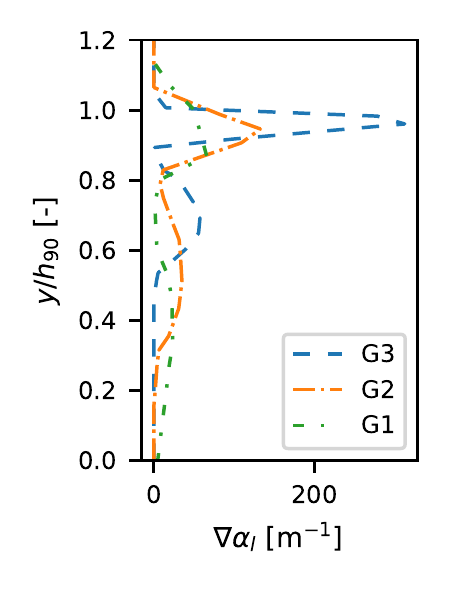}

      \caption{Left: Typical distribution pattern of an unrestricted $S_g$. Right: Typical profiles of the gradient of $\alpha_l$. }
      \label{fig:sg_and_grad_alpha}
\end{figure}

Before discussing the formulation of $\delta_{fs}$, it is necessary to consider what kind of restriction of $S_g$ in space is needed in the case of spillway flow.
A typical distribution of an unrestricted $S_g$ is shown in the left plot of Figure~\ref{fig:sg_and_grad_alpha}.
The first thick yellow line is located right below the interface and represents the region where the entrainment can be expected to take place.
However, a discontinuous strip of non-zero values is also observed close to the psuedobottom along with more or less randomly distributed points of activation in the corners of the steps.
Physically, no entrainment can occur in these regions, and the $\delta_{fs}$ function should filter them out.
Note that the spatial separation between the correct and non-physical regions of source term activation is not large, which explains why defining $\delta_{fs}$ in a universal way that fits all flow conditions and numerical settings is not trivial.

\begin{figure}[htp]
	\vspace{-8pt}
	\centering
	\includegraphics[scale=1]{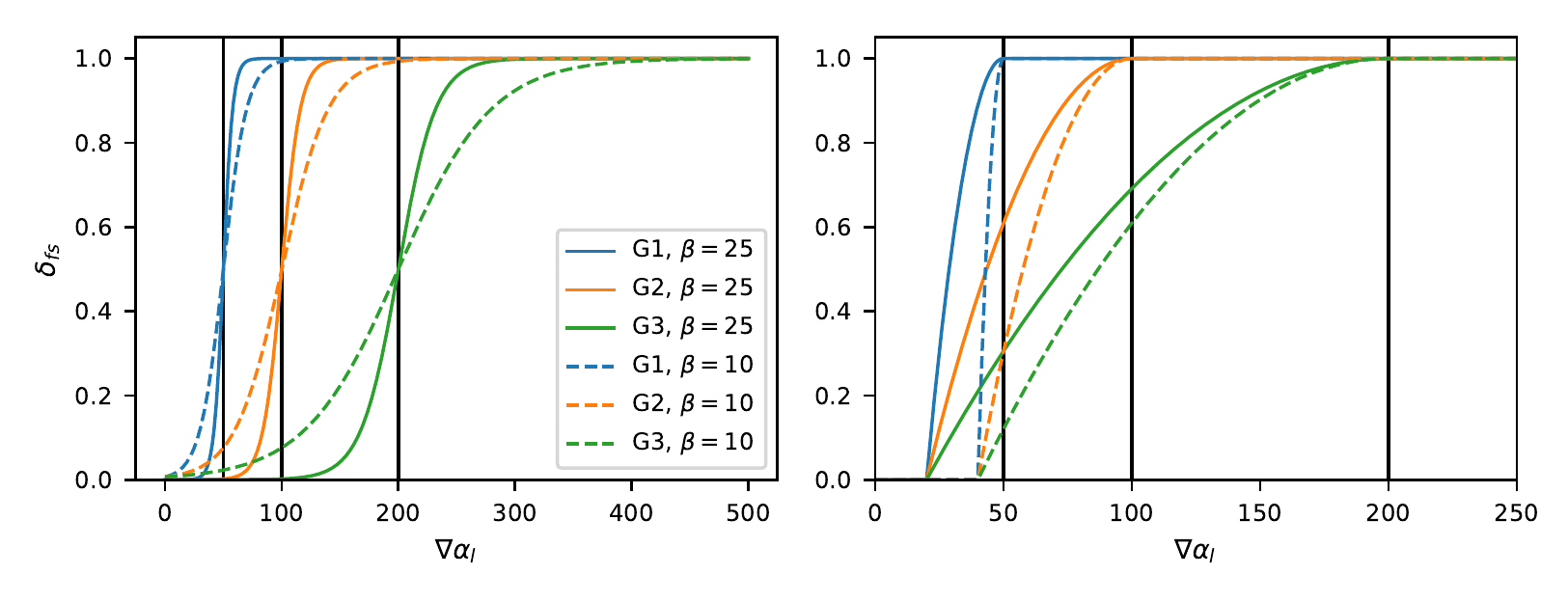}
	\caption{Surface indicator functions  $\delta_{fs}$.}
	\label{fig:delta_fs}
\end{figure}

To arrive to a better formulation for $\delta_{fs}$, it is important to clearly understand the deficiencies of the original definition, see Eq.~\eqref{eq:aifdeltafs}.
With $\nabla \alpha_{l,cr}$ set to $1/(4\Delta x)$, the distribution of $\delta_{fs}$ over $\nabla \alpha_{l}$ depends on two quantities: $\beta$ and  $\Delta x$.
It is instructive to see how $\delta_{fs}$ changes shape when the values of these parameters are changed.
In the left plot of Figure~\ref{fig:delta_fs}, $\delta_{fs}(\nabla \alpha_l)$ is shown for $\Delta x$ values corresponding to grids G1-G3, and the two values of $\beta$ considered in the sensitivity study in Section~\ref{sec:aifbeta}.
The tanh function defining the transition region of $\delta_{fs}$ from 0 to 1 is centred at $\nabla \alpha_{l,cr}$, shown in the figure with black vertical lines.
As the grid is refined, this location is shifted to the right, and for larger values of $\beta$, the tanh only spans a limited range of high $\nabla \alpha_l$ values.
On the other hand, for smaller $\beta$, the tanh becomes so wide that
$\delta_{fs}$ remains non-zero everywhere.

This behaviour of $\delta_{fs}$ should be related to how $\nabla \alpha_l$ is typically distributed across $y$, see the right plot in Figure~\ref{fig:sg_and_grad_alpha}.
Note that the high values of $\nabla \alpha_l$ are always restricted to to a relatively thin region close to the interface.
Consequently, for a large $\beta$, for example $\beta = 100$ as proposed in~\cite{Lopes2017}, $\delta_{fs}$ will be restricted to an increasingly smaller region in space when the grid gets refined.
This explains why the diminishing effect of the entrainment with grid refinement observed in Section~\ref{sec:gridsens}.

As mentioned above, it is easy to make the $\delta_{fs}$ function less restrictive by lowering $\beta$.
However, the issue here is that the $\beta$ and the effective cut-off value in terms of $\nabla \alpha_{l}$ are not intuitively related.
This is problematic given that the margin of error is quite small, as discussed above in relation to the typical distribution of $S_g$.

Alternative options for the expression of $\delta_{fs}$ were explored, including
a parabola based function, and a step-shaped function defined purely based on
the distance from a defined interface.   
For both alternatives, the idea was to set the value of the critical gradient of
$\alpha_l$ relatively tight, to capture the upper peak in
the gradient plot in 
Figure~\ref{fig:sg_and_grad_alpha}, and then expand its prevalence away from
the these locations according to an appropriate function or logic.
In the distance based alternative, $\delta_{fs}$ was set to 0 or 1 for a
particular cell, depending on its distance from the defined interface. If this
distance was less then the interface thickness, $\phi_{ent}$, $\delta_{fs}$ was
set to 1, and otherwise its value was set to 0. 
However, this led to step-shaped profiles of $\alpha_l$, where too much air
was entrained below the interface, and indicated a need to apply
some functionality to reduce the effects of the source term as the gradient of
$\alpha_l$ is reduced below its critical value. 

Acknowledging the above, a parabola based function was 
considered. The possibility to define $\nabla \alpha_{cr}$ as a top point of the
function, and to define a cut-off value for the gradient as an intersection
point, was viewed as beneficial features of this function in the current
setting. The latter leading to the possibility of avoiding the long tail in the
tanh-function, with the 
corresponding uncontrollable potential for generation of none-zero values of
$S_g$ within the steps.
The parabola-based $\delta_{fs}$ formulation is proposed as

\begin{equation}
  \delta_{fs}(\nabla \alpha_l) =
  \begin{cases}
    \text{Pos}\left(-\frac{1}{4d} \big(|\nabla \alpha_l| - |\nabla
  \alpha_{l,cr}|\big)^2 +1\right)  & \text{if $\nabla \alpha_l < |\nabla \alpha_{l,cr}|$} \\
    1 & \text{otherwise}. \\
  \end{cases}
  \label{eq:deltafspara}
\end{equation}
Here, $d$ refers to the distance from the vertex of the parabola to its focus, which can be computed as
\[
d = 0.25 \left(|\nabla \alpha_{l,cr}| -
|\nabla \alpha_{l,cut}|\right)^2,
\]

\noindent
where $\nabla \alpha_{l,cut}$ is an input parameter explicitly defining the lowest $\nabla \alpha_{l}$ for which the source term may assume non-zero values.
The proposed $\delta_{fs}$, computed for grids G1-G3 and two different values of $\nabla \alpha_{l,cut}$, is shown in the right plot of Figure~\ref{fig:delta_fs}.
The non-zero values of the function are always fixed to the interval $[\nabla \alpha_{l,cut}, \nabla \alpha_{l,cr}]$, which expands upon grid refinement.
Unfortunately, due to this expansion, even with this new $\delta_{fs}$, the region of non-zero $S_g$ values shrinks as the grid is refined.
However, the process is slowed, since $\delta_{fs}$ is left to be non-zero at lower $\nabla \alpha_l$.

During initial testing, it was observed that the parabola-based $\delta_{fs}$ was very effective at filtering out the sporadic source term activation in the corner of the steps, while preserving the region where entrainment is expected.
However, the secondary strip of non-zero $S_g$ values close to the psuedobottom (see left plot in Figure~\ref{fig:sg_and_grad_alpha}) would sometimes still be left unfiltered.
This is likely related to the secondary peak in $\nabla \alpha_l$.
To rectify this, the parabolic surface indicator function in
Eq.~(\ref{eq:deltafspara}) is combined with a distance based approach like the
one outlined above.

In details, the values of $\delta_{fs}$ computed according to Eq.~\ref{eq:deltafspara} are additionally manipulated as follows.
First, the cells in which $\delta_{fs} \geq 0.9$ are selected.
These should lie near the interface and are therefore likely to belong to the region where $S_g$ should be activated.
For the remaining cells, the distance to the nearest cells with $\delta_{fs} \geq 0.9$ is computed.
If this distance exceeds the interface thickness, $\phi_{ent}$, the value
of $\delta_{fs}$ in the cell is set to 0.
Overall, this combined
formulation gave improvements compared to the original formulation based on
tanh. It is noted that even without the distance cut-off modification, improved
results with the parabolic $\delta_{fs}$ could be achieved, and the $\delta_{fs}
\geq 0.9$ criterion can give false positives.

Finally, making both $\nabla \alpha_{l,cr}$ and $\nabla \alpha_{l,cut}$ grid independent
parameters has briefly been tested, but abandoned due to the sensitivity of the
results to the chosen values.
In general, our extensive efforts and experimenting with $\delta_{fs}$ formulations and other simulation settings showed that constructing a robust and accurate $\delta_{fs}$ is very challenging.
The sensitivity of the results to any changes in the simulation parameters or flow conditions tends to be very strong.
Nevertheless, as shown in the Section~\ref{sec:newres}, the proposed $\delta_{fs}$ does represent an improvement with respect to prior art.

\subsection{Modelling air propagation into the corners of the steps}
\label{subsec:extradiff}

By definition, the employed entrainment model is meant to account for aeration occurring close to the free surface, 
within some layer of thickness $\phi_{ent}$.
However, experimental results clearly show that in the case of the stepped spillway, air penetrates all the way down to the surface of the steps, see e.g.~the experimental profiles in Figure~\ref{fig:fsComp_alpha_dx005}.
The physical mechanism through which this occurs is described by \citet{Pfister2011}.
Inspection of video recordings from their experiments reveals the occasional generation of air troughs that extend from the surface into the bulk flow.
These troughs penetrate deep enough to hit the step edges, and when they do, the air is distributed into the steps.

Capturing this intrinsically transient process in a steady state model is not
straightforward.
Here, we consider using a somewhat ad-hoc approach, taking advantage of the diffusion term in the $\alpha_g$ equation~\eqref{eq:alphag}, $ \nabla \cdot (\nu_t \nabla \alpha_g)$.
Recall that in Section~\ref{sec:aifur} it was shown that the effect of this term on the solution is essentially negligible.
However, the effect can be easily amplified by pre-multiplying it with some constant $C_t$
.
The increased diffusion of $\alpha_g$ will then lead to air being redistributed more evenly across $y$, and consequently result in stronger aeration closer to the pseudobottom.

The difficulty lies in the choice of the value of $C_t$ since there is no clear physical analogy between the modelled phenomenon and diffusion.
In light of this, it was attempted to search for a suitable value through experimentation, to see whether one leading to improved results across all the considered Froude numbers could be found.
As a result, $C_t = 150$ was selected.

\subsection{Inception point prediction}
\label{subsec:sgact}
Even if AIF predicts the aeration onset correctly when appropriate values for $k_c$ and $u_c$ are supplied, the inception point estimation using this method completely depends on user input.
As shown in Section \ref{sec:aiffscomp}, the appropriate critical value $k_c$
depends on the flow conditions.
An alternative approach, found in the work of \citet{hirt2003modeling}, is to directly consider the balance between the energy of turbulent motion and that of gravity and surface tension.
Defining
\begin{align}
  \label{eq:pt}
&  P_t = \rho k, \\
  \label{eq:pd}
&  P_d = \rho |g| a + \frac{\sigma}{a},
\end{align}
the source term $S_g$ is activated only in cells where $P_t > P_d$.
This way the inception point prediction requires no user input.
However, it completely relies on the correct prediction of $k$ and $a$ by the turbulence model.

\section{Simulations with the improved model}
\label{sec:newres}
This section is dedicated to evaluating the effects of the model improvements proposed above.
To that end, a new solver incorporating these changes has been implemented.
Reflecting the focus on stepped spillway simulations, the solver is called
\ttt{spillwayFlow} \footnote{https://github.com/siljekre/spillwayFlow}, which is
abbreviated to SPF below.
The robustness of the model with respect to grid resolution is evaluated in Section~\ref{subsec:spf_grid}.
Results from application of the model to spillway flow at all four considered Froude numbers are presented Section \ref{subsec:spf_fs}.
The proposed criterion for inception point location is tested separately, in Section~\ref{subsec:spf_icpt}.

\subsection{Grid sensitivity}
\label{subsec:spf_grid}

\begin{figure}[htp]
   \vspace{-7pt}
   \centering
   \subcaptionbox{Surface elevation plot, $h_{90}$-surface.\label{subfig:spfh90_nodiff_fs27_aif_grid}}
        {\includegraphics[scale=1,trim={0.4cm 10 0.2cm
              3},clip]{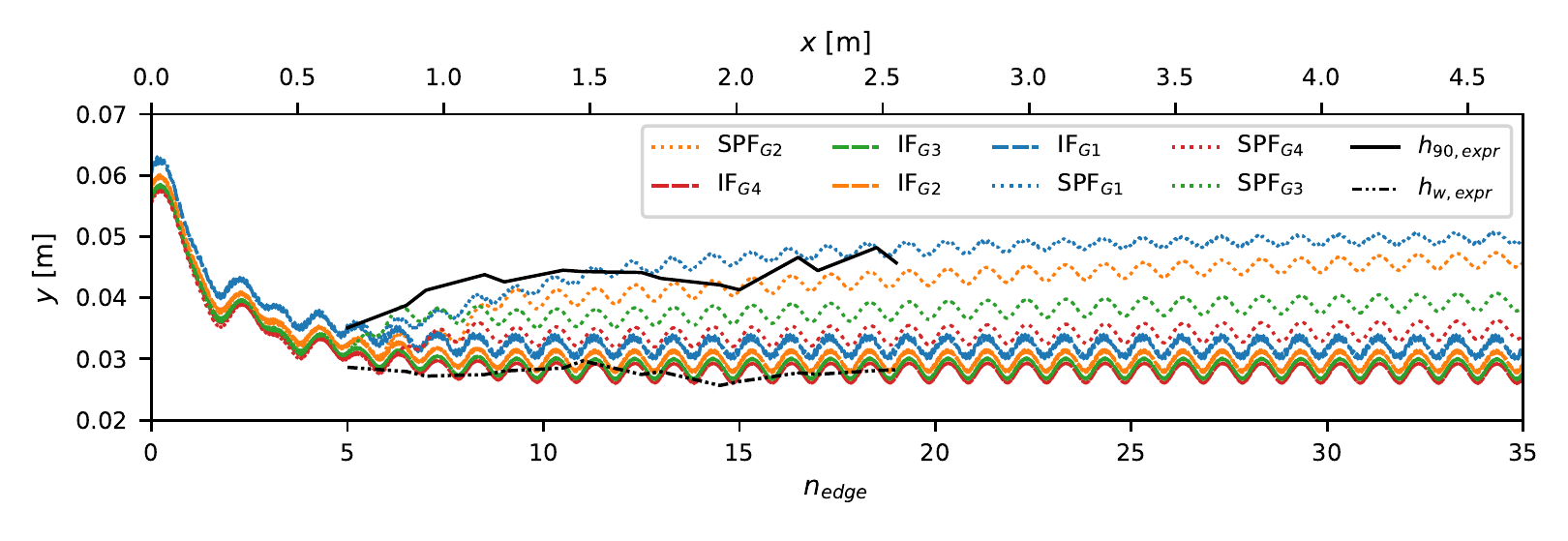}}

      \subcaptionbox{Step 7\label{subfig:spfgridComp_nodiff_alpha_M7}}
        {\includegraphics[scale=1,trim={0.1cm 10 0.2cm 3}]{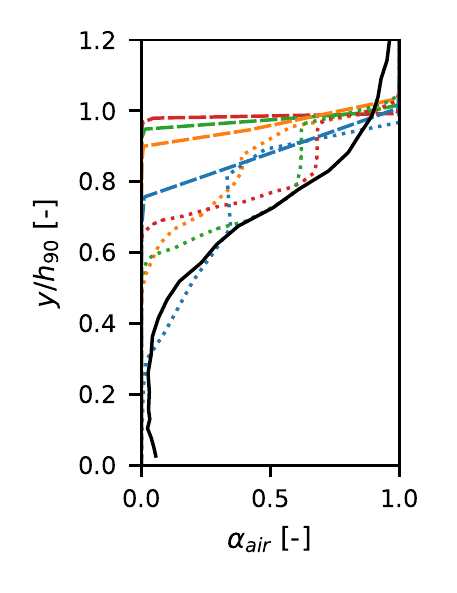}}
      \subcaptionbox{Step 11\label{subfig:spfgridComp_nodiff_alpha_M11}}
        {\includegraphics[scale=1,trim={0.7cm 10 0.2cm 3},clip]{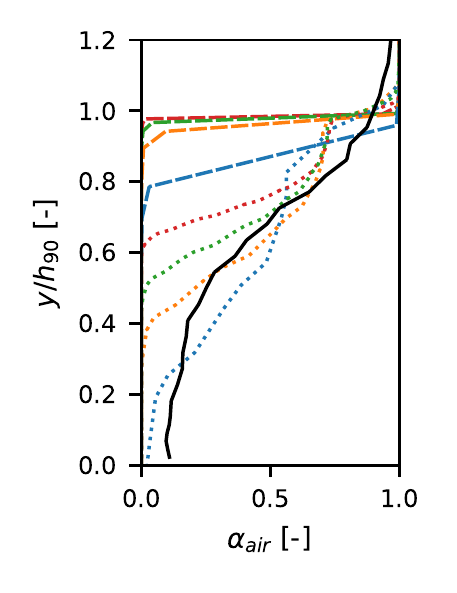}}
      \subcaptionbox{Step 15\label{subfig:spfgridComp_nodiff_alpha_M15}}
        {\includegraphics[scale=1,trim={0.7cm 10 0.2cm
              3},clip]{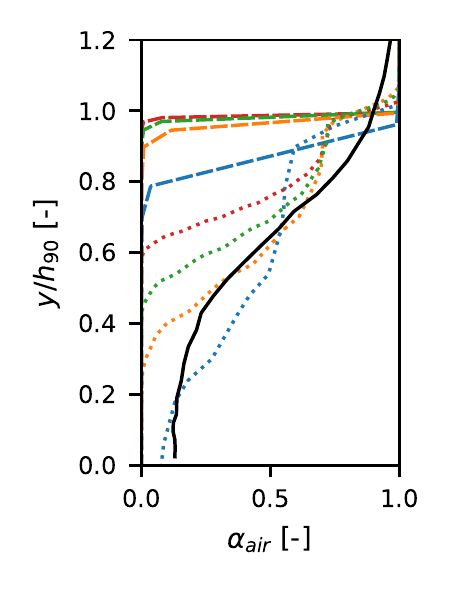}}
      \subcaptionbox{Step 19\label{subfig:spfgridComp_nodiff_alpha_M19}}
        {\includegraphics[scale=1,trim={0.7cm 10 0
              3},clip]{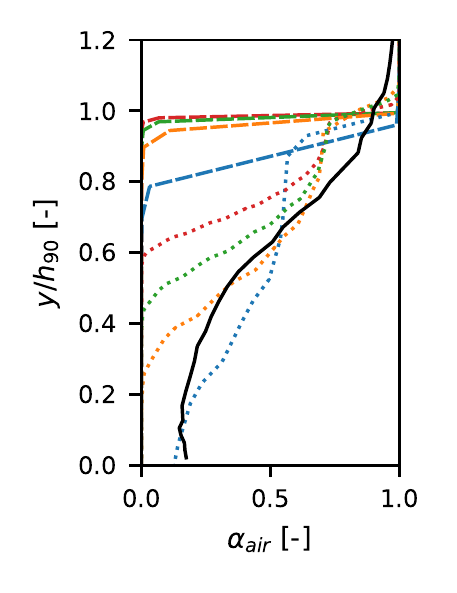}}


        \caption{SPF simulated with no $\alpha_g$-diffusion ($C_t=0$) and
          \textsf{F\textsubscript{s}}=2.7 at different grids 
          (G1-G4) compared to physical model results by
          \citet{Bung2011}. Figure~\ref{subfig:spfh90_nodiff_fs27_aif_grid} shows the
          surface elevation, Figures
          \ref{subfig:spfgridComp_nodiff_alpha_M7}-\ref{subfig:spfgridComp_nodiff_alpha_M19},
          the void fraction profiles at different steps.
      }\label{fig:spfgridComp_nodiff_fs27}
\end{figure}

Here the new model is put to the same grid sensitivity analysis as presented in Section~\ref{sec:gridsens} for AIF.
To simplify the analysis, the new inception point prediction approach is not
employed, and the original criterion based on $k_c$ is used
instead.  
Simulation results for the \textsf{F\textsubscript{s}}=2.7 case obtained on grids G1-G4 are shown in Figure~\ref{fig:spfgridComp_nodiff_fs27}.
Here, $C_t = 0$ and the diffusion term in the $\alpha_g$ equation is thus inactive.
As anticipated, the results still depend on the grid, and the general trend is convergence towards the IF solution.
Note that in the $\alpha_{air}$ profiles, the reduction of aeration manifests itself predominately at $y/h_{90} \lessapprox 0.6$.
Closer to $h_{90}$ results remain acceptable even on the G4 grid.
Related to the absence of air in the lower parts, the predictions of $h_{90}$
itself are more sensitive, and unfortunately on the finer grids the accuracy is poor. 
Nevertheless, compared to the original AIF results (see
Figure~\ref{fig:aifgridComp}) the robustness of the model is
improved.

\begin{figure}[htp]
   \vspace{-7pt}
   \centering
   \subcaptionbox{Surface elevation plot, $h_{90}$-surface.\label{subfig:spfh90_fs27_aif_grid}}
        {\includegraphics[scale=1,trim={0.4cm 10 0.2cm
              3},clip]{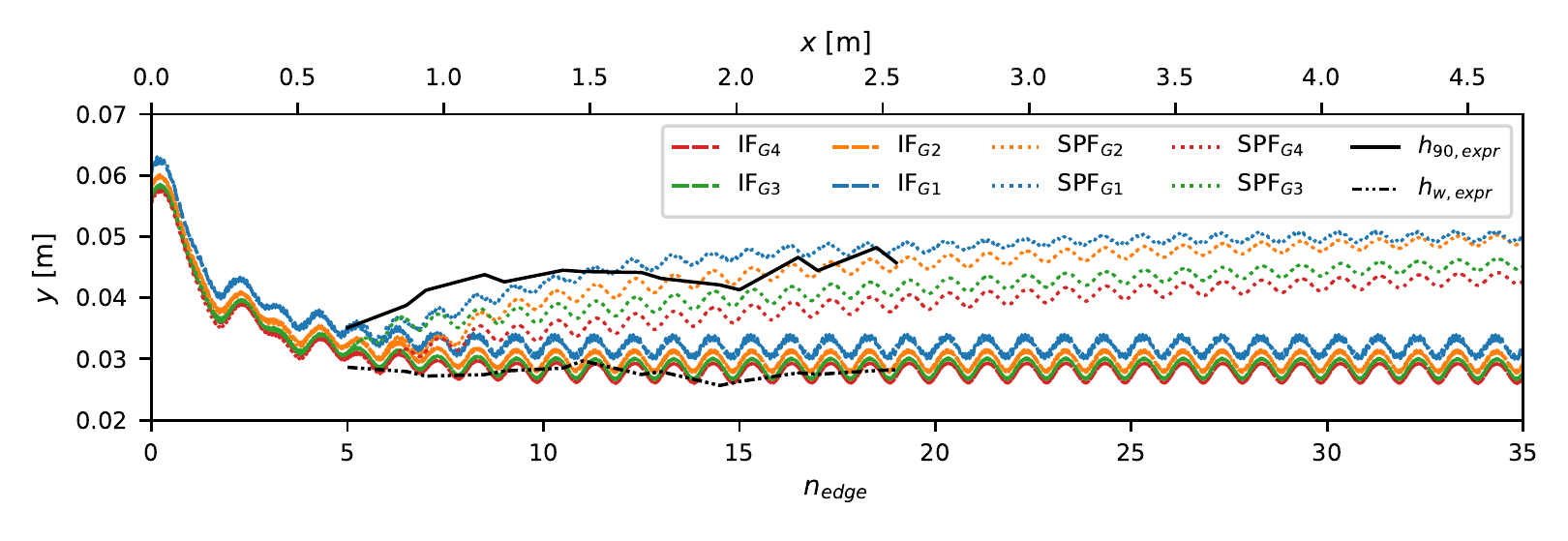}}

      \subcaptionbox{Step 7\label{subfig:spfgridComp_alpha_M7}}
        {\includegraphics[scale=1,trim={0.1cm 10 0.2cm 3}]{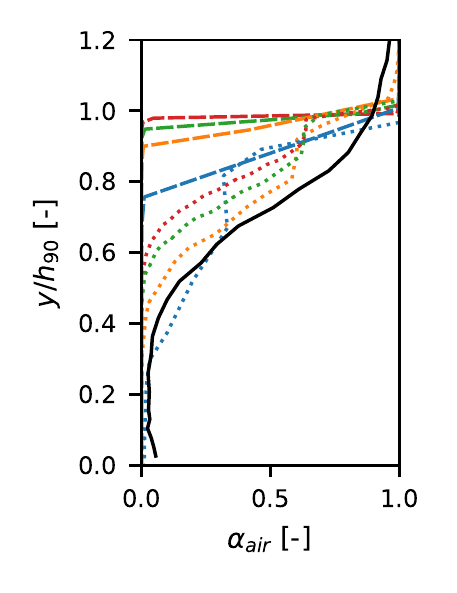}}
      \subcaptionbox{Step 11\label{subfig:spfgridComp_alpha_M11}}
        {\includegraphics[scale=1,trim={0.7cm 10 0.2cm 3},clip]{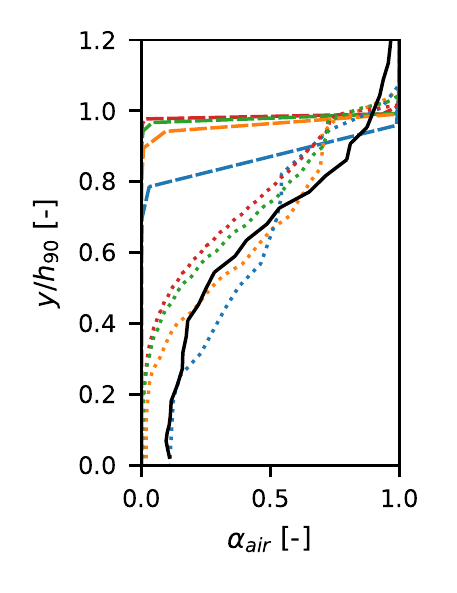}}
      \subcaptionbox{Step 15\label{subfig:spfgridComp_alpha_M15}}
        {\includegraphics[scale=1,trim={0.7cm 10 0.2cm
              3},clip]{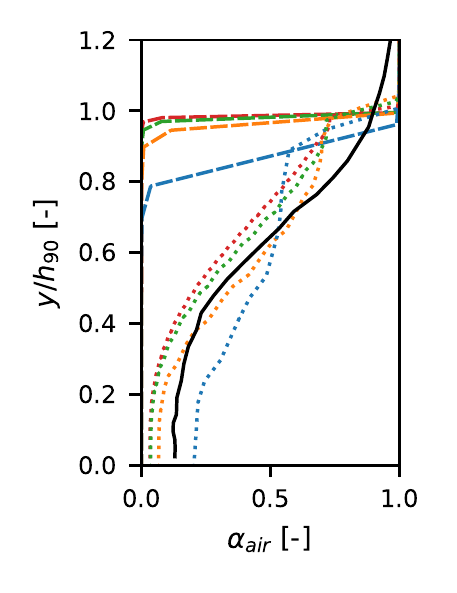}}
      \subcaptionbox{Step 19\label{subfig:spfgridComp_alpha_M19}}
        {\includegraphics[scale=1,trim={0.7cm 10 0
              3},clip]{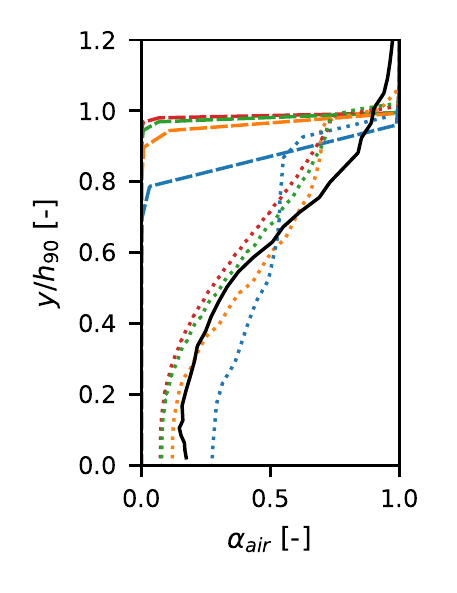}}

        \caption{SPF simulated with $\alpha_g$-diffusion ($C_t=150$) and
          \textsf{F\textsubscript{s}}=2.7 at different grids 
          (G1-G4) compared to physical model results by
          \citet{Bung2011}. Figure~\ref{subfig:spfh90_fs27_aif_grid} shows the
          surface elevation, Figures
          \ref{subfig:spfgridComp_alpha_M7}-\ref{subfig:spfgridComp_alpha_M19},
          the void fraction profiles at different steps in the developing
          region. 
      }\label{fig:spfgridComp_fs27}
\end{figure}

Figure~\ref{fig:spfgridComp_fs27} shows the results obtained with $C_t = 150$.
As expected, a comparatively more even distribution of $\alpha_{air}$ across $y$ is achieved, in particular for the simulations on denser grids.
Furthermore, a very clear improvement in the robustness of the model is evident, with much more similar results obtained on all four grids.

\subsection{Results for different Froude numbers}
\label{subsec:spf_fs}
Now, the results obtained with the new model are presented for all four considered values of the Froude number.
The simulations were run on the densest mesh, G4.
As in the section above, the new criterion for locating the inception point is not used here, and instead $k_c$ is adjusted for each case in order to match the location in the experimental data.
For completeness, profiles corresponding to both $C_t = 0$ and $C_t = 150$ are shown in all the figures.
For comparison, they also include results from AIF simulations on the G1 grid, and also from IF simulations on the G4 grid.

\begin{figure}[htp]
   \vspace{-8pt}
   \centering
      \subcaptionbox{\textsf{F\textsubscript{s}}=2.7\label{subfig:fs27_alpha_final}}
        {\includegraphics[scale=1,trim={0.0 10 0 0},clip]{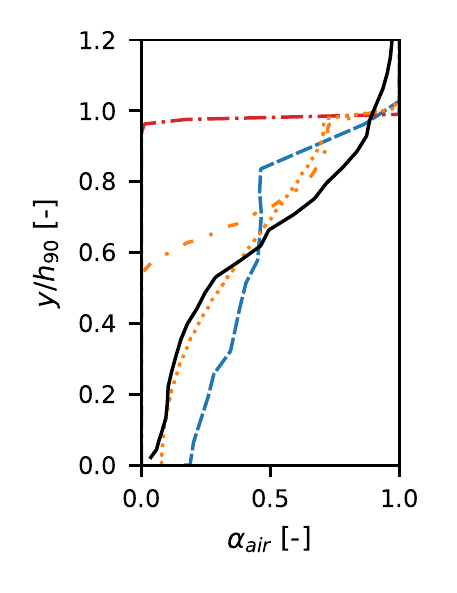}}
      \subcaptionbox{\textsf{F\textsubscript{s}}=4.6\label{subfig:fs46_alpha_final}}
                    {\includegraphics[scale=1,trim={0.8cm 10 0 0},clip]{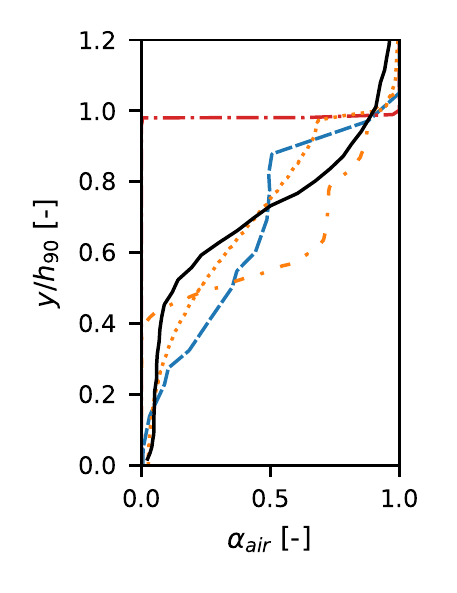}}
      \subcaptionbox{\textsf{F\textsubscript{s}}=8.28\label{subfig:fs828_alpha_final}}
        {\includegraphics[scale=1,trim={0.8cm 10 0
              0},clip]{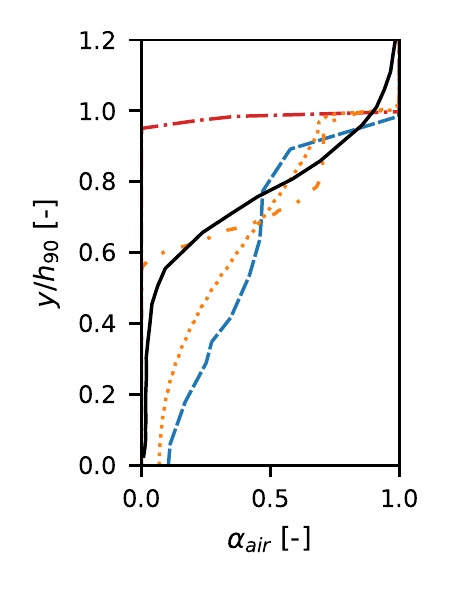}}
      \subcaptionbox{\textsf{F\textsubscript{s}}=13\label{subfig:fs13_alpha_final}}
        {\includegraphics[scale=1,trim={0.8cm 10 0
              0},clip]{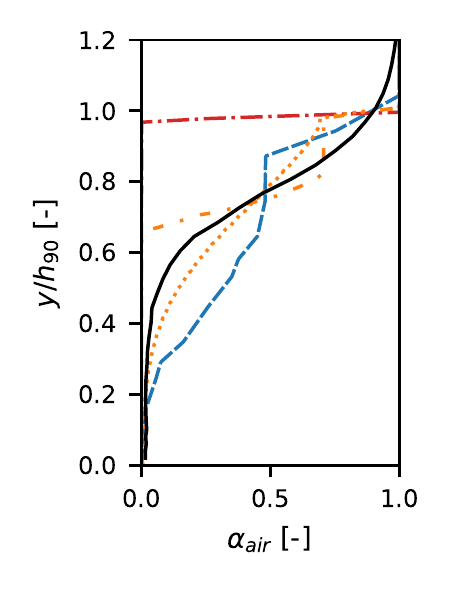}}
        \\
        \caption{Vertical void fraction profiles for uniform flow
          conditions. Spillway flows at different Froude numbers.
          SPF simualtions (G4) compared to IF (G4), AIF (G1) and experimental
          results by \citet{Bung2011}. All plots are showing profiles at step
          edges.}\label{fig:fsComp_alpha_final}
\end{figure}

The $\alpha_{air}$ profiles are discussed first, see Figure~\ref{fig:fsComp_alpha_final}.
Qualitatively, the same behaviour with respect to $C_t$ is observed for all \textsf{F\textsubscript{s}}.
With $C_t=0$ the distribution of $\alpha_{air}$ across $y$ is close to step-wise, with decent agreement with experimental data for $y/h_{90} \gtrapprox 0.7$.
When $C_t=150$, the profiles are smoothed out, which generally increases the accuracy.
The exception is the \textsf{F\textsubscript{s}}=8.28 case, for which the air volume fraction at low $y/h_{90}$ becomes excessive.
Yet even for this case the agreement is better than what could be achieved with AIF.

\begin{figure}[htp]
  \vspace{-20pt}
   \centering
    \begin{subfigure}[b]{\textwidth}
        \includegraphics[scale=1,trim={0 20 0 0},clip]{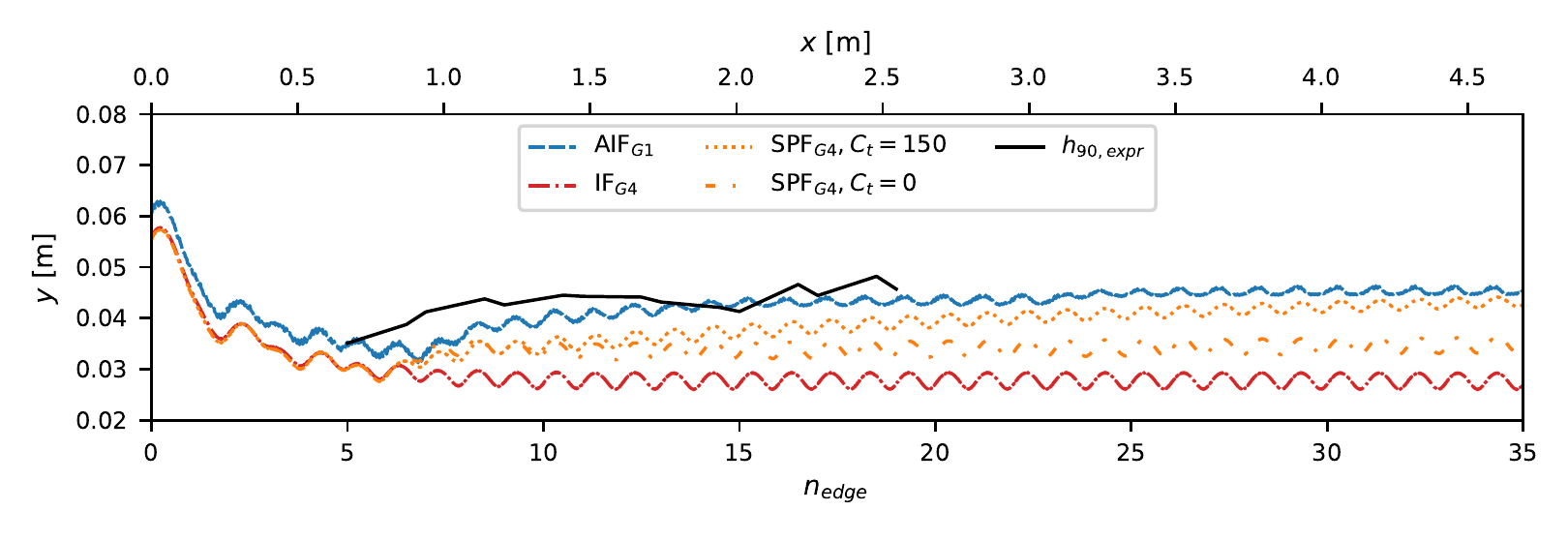}
        \caption{\textsf{F\textsubscript{s}}=2.7}
        \label{subfig:h90_fs27_final}
    \end{subfigure}
    \par\medskip
    \begin{subfigure}[b]{\textwidth}
        \includegraphics[scale=1,trim={0 20 0 20},clip]{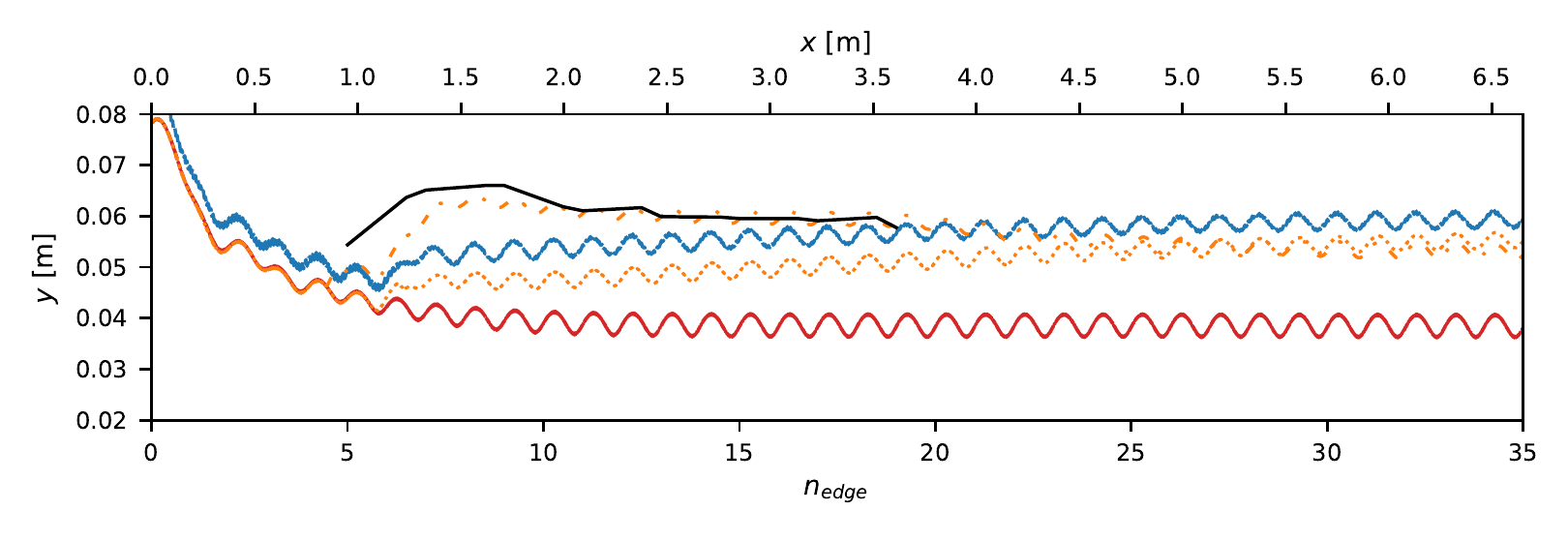}
        \caption{\textsf{F\textsubscript{s}}=4.6}
        \label{subfig:h90_fs46_final}
    \end{subfigure}
    \par\medskip
    \begin{subfigure}[b]{\textwidth}
        \includegraphics[scale=1,trim={0 20 0 20},clip]{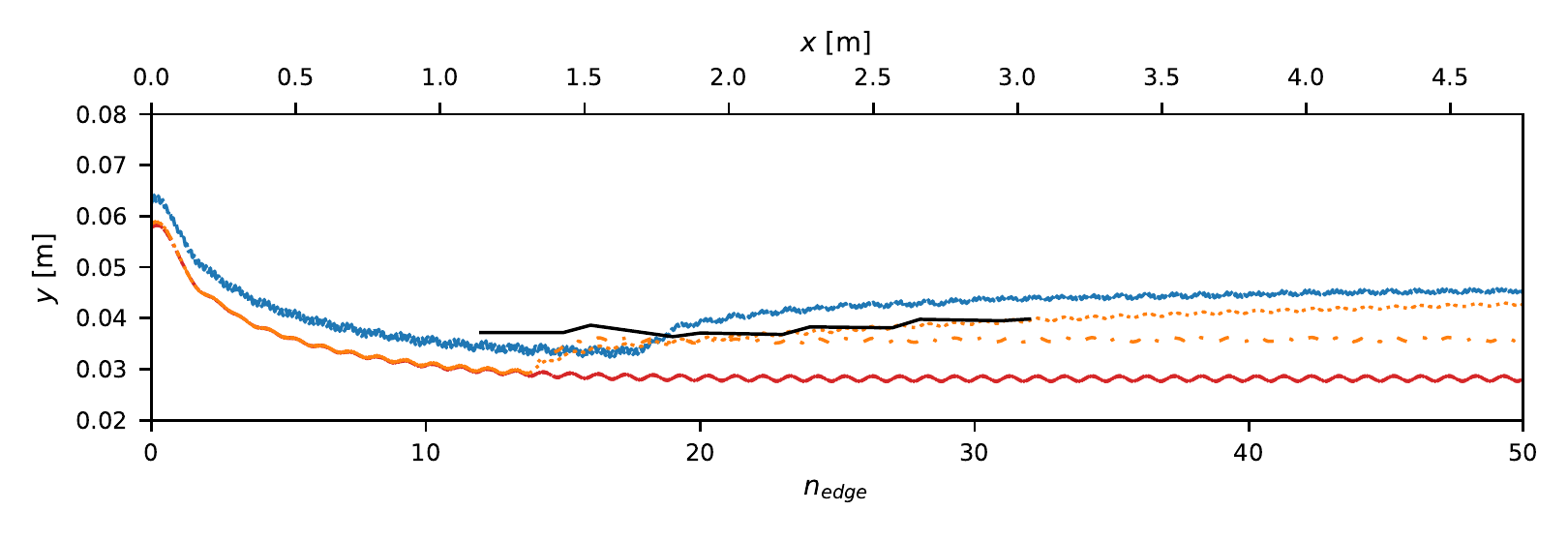}
        \caption{\textsf{F\textsubscript{s}}=8.28}
        \label{subfig:h90_fs82_final}
    \end{subfigure}
    \par\medskip
        \begin{subfigure}[b]{\textwidth}
        \includegraphics[scale=1,trim={0 10 0
            20},clip]{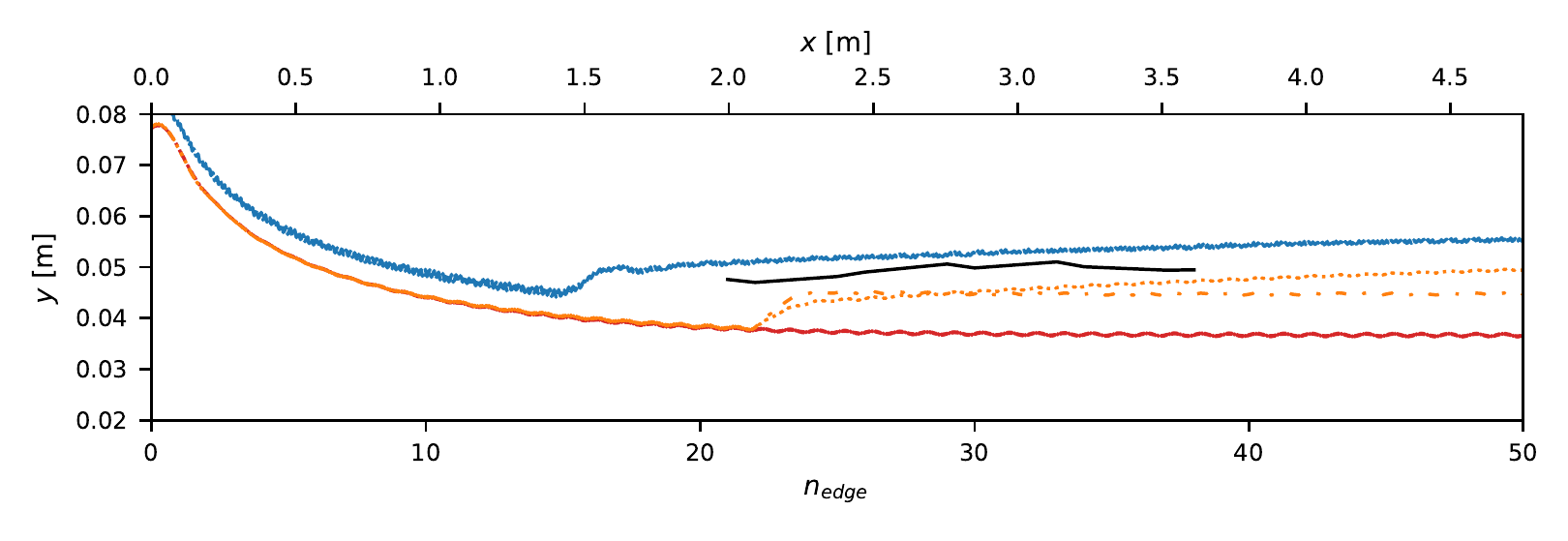}
        \caption{\textsf{F\textsubscript{s}}=13}
        \label{subfig:h90_fs13_final}
    \end{subfigure}
    \caption{Surface elevation plots, $h_{90}$. Spillway flows at different
      Froude numbers. SPF simualtions (G4) compared to IF (G4), AIF (G1) and
      experimental results by \citet{Bung2011}.  
    }\label{fig:h90_fscomp_final}
\end{figure}

The surface elevation plots are shown in Figure~\ref{fig:h90_fscomp_final}.
Overall, $C_t = 150$ leads to better results, which agree well with the experimental data in the uniform flow region.
An interesting exception is the \textsf{F\textsubscript{s}}=4.6, for which $C_t$ barely has influence on $h_{90}$ in the uniform flow region, whereas in the developing region $C_t = 0$ leads to very good agreement with the experiment.
However, it is unlikely that this is explained by any fundamental property of the flow or the model.
Compared to AIF, the accuracy of the new model is generally on par.
AIF curves are marginally closer to experimental data for the two lower \textsf{F\textsubscript{s}}, and the other way around for the two higher \textsf{F\textsubscript{s}}.

\subsection{Inception point analysis}
\label{subsec:spf_icpt}
In this section, the source term activation criterion presented in Section~\ref{subsec:sgact} is tested.
Recall that the inception point location is determined from the flow, and depends heavily on the employed turbulence model.
Here we show results from simulations using four models: the $k$-$\omega$ SST and realisable $k$-$\eps$ from the standard OpenFOAM library, and their respective counterparts in the library by \citet{Fan2020}, in which the density gradient is properly accounted for in the transport equations.
The latter are referred to by adding \texttt{varRho} to the name of the
model.
Simulations using standard turbulence modelling and the $k_c$-criteria
purposed by \citet{Lopes2017} are added as reference.
The simulations are performed on the G3 grid.

\begin{figure}[htp]
  \vspace{-10pt}
  \centering
  \begin{subfigure}[b]{0.49\textwidth}
    \includegraphics[scale=1,trim={0.28cm 29 0.2cm
        0},clip]{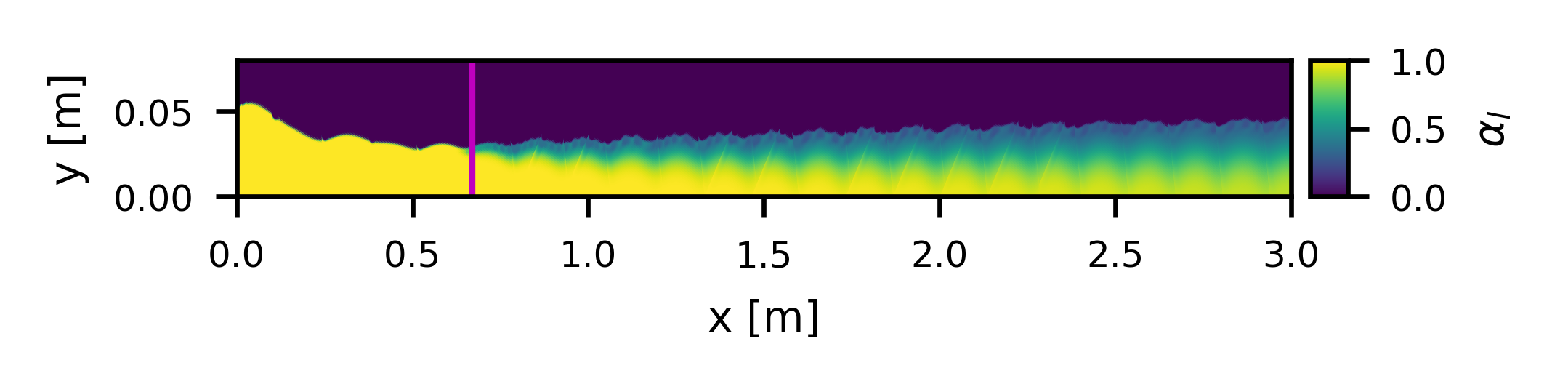}
    \\
    \includegraphics[scale=1,trim={0.28cm 29 0.2cm
        0},clip]{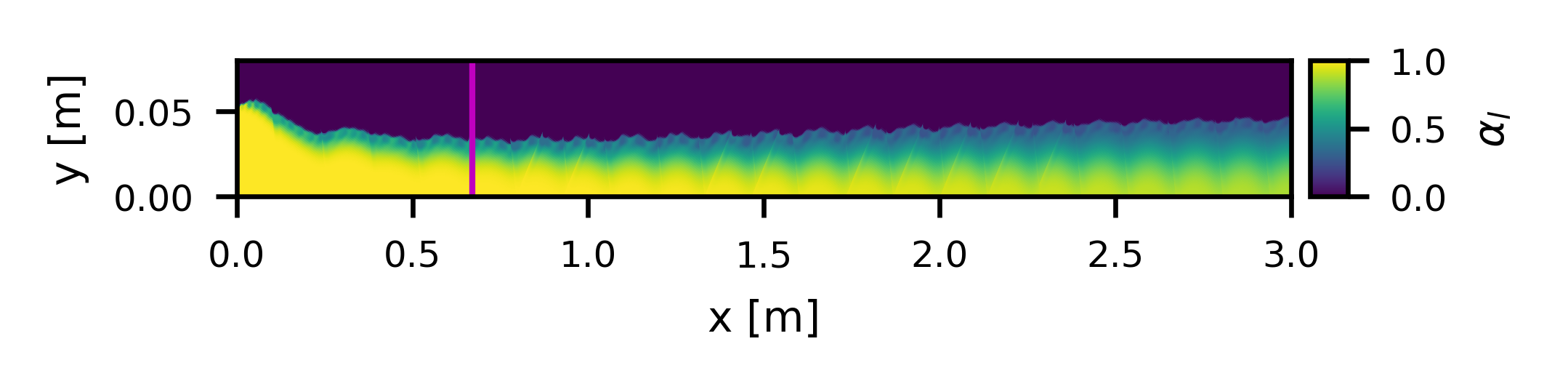}
    \\
    \includegraphics[scale=1,trim={0.28cm 8 0.2cm
        0},clip]{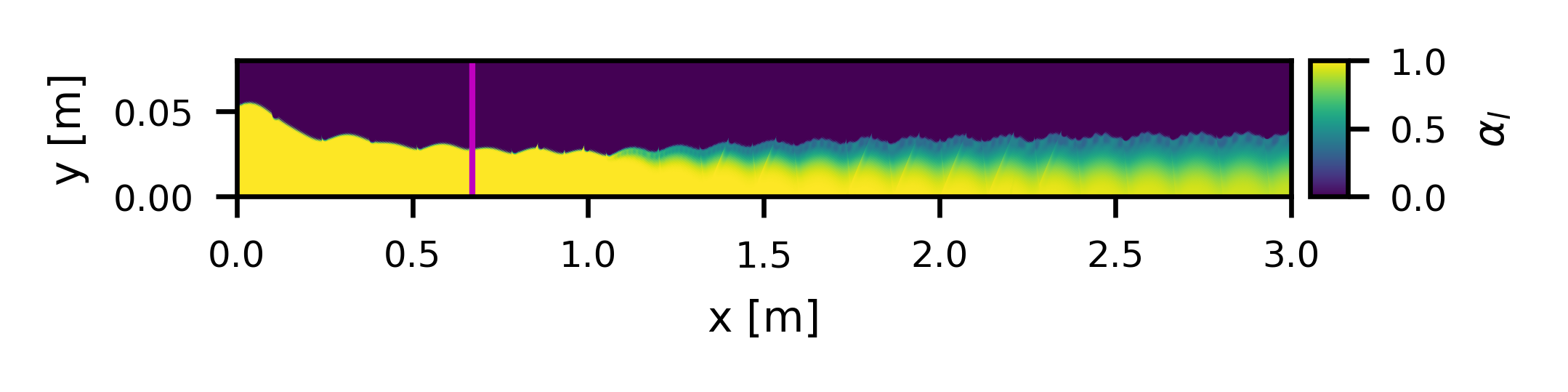}
    \caption{\textsf{F\textsubscript{s}}=2.7 -- realisable $k$-$\eps$}
    \label{subfig:icpt_fs27}
  \end{subfigure}
  \begin{subfigure}[b]{0.49\textwidth}
    \includegraphics[scale=1,trim={0.6cm 29 0.2cm
        0},clip]{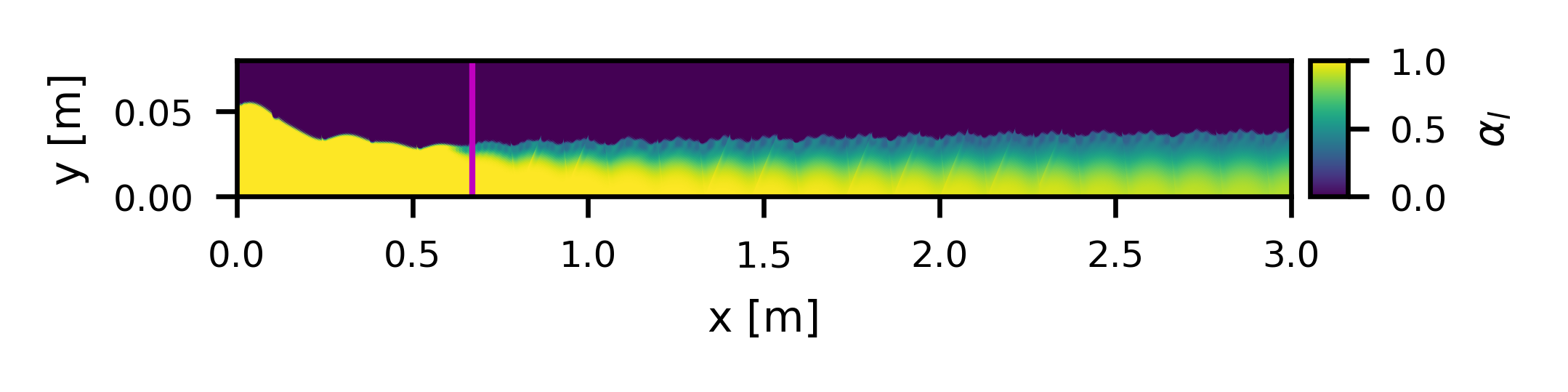}
        \\
    \includegraphics[scale=1,trim={0.6cm 29 0.2cm
        0},clip]{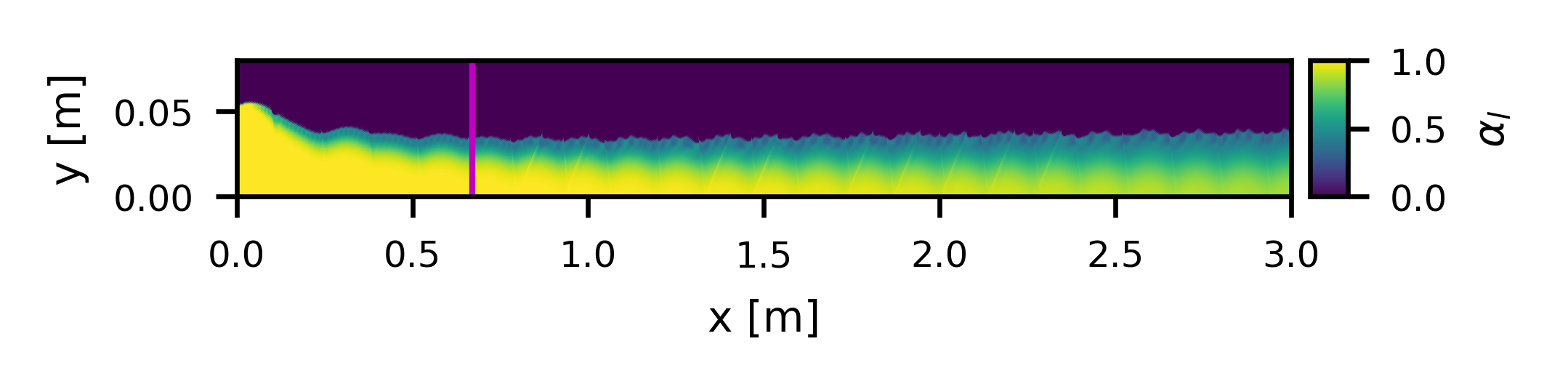}
        \\
    \includegraphics[scale=1,trim={0.6cm 8 0.2cm
        0},clip]{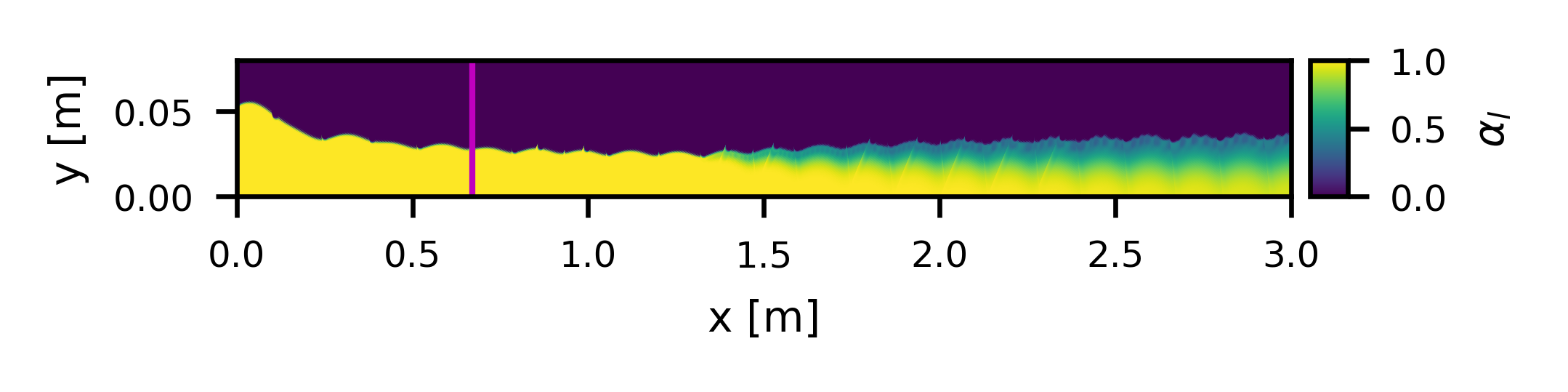}
        \caption{\textsf{F\textsubscript{s}}=2.7 -- $k$-$\omega$ SST}
        \label{subfig:icpt_fs27}
  \end{subfigure}
  \vspace{-3pt}
  \begin{subfigure}[b]{0.49\textwidth}
    \raisebox{-\height}
    {\includegraphics[scale=1,trim={0.28cm 29 0.2cm
        0},clip]{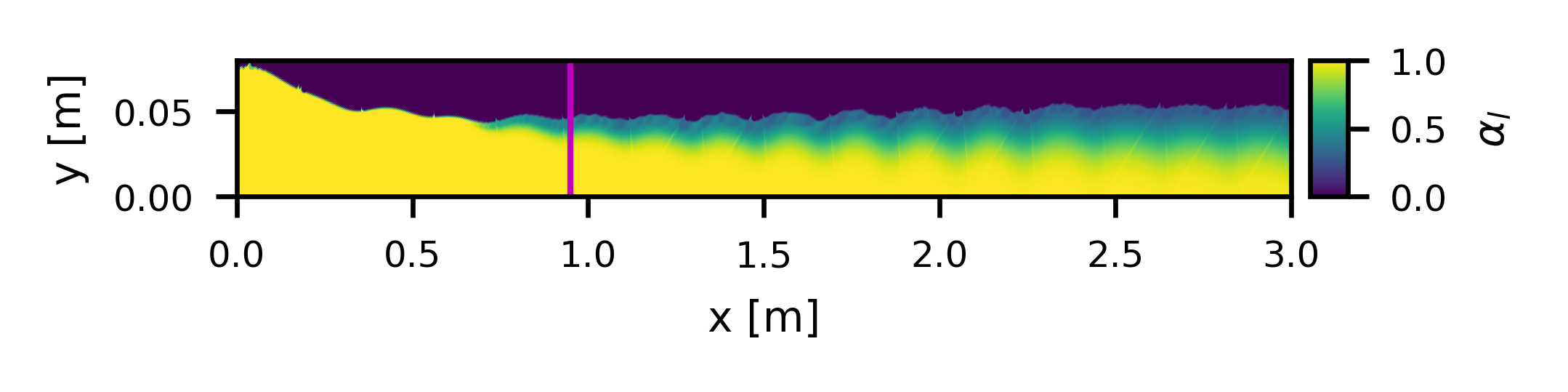}}
    \\    
    {\includegraphics[scale=1,trim={0.28cm 29 0.2cm
        0},clip]{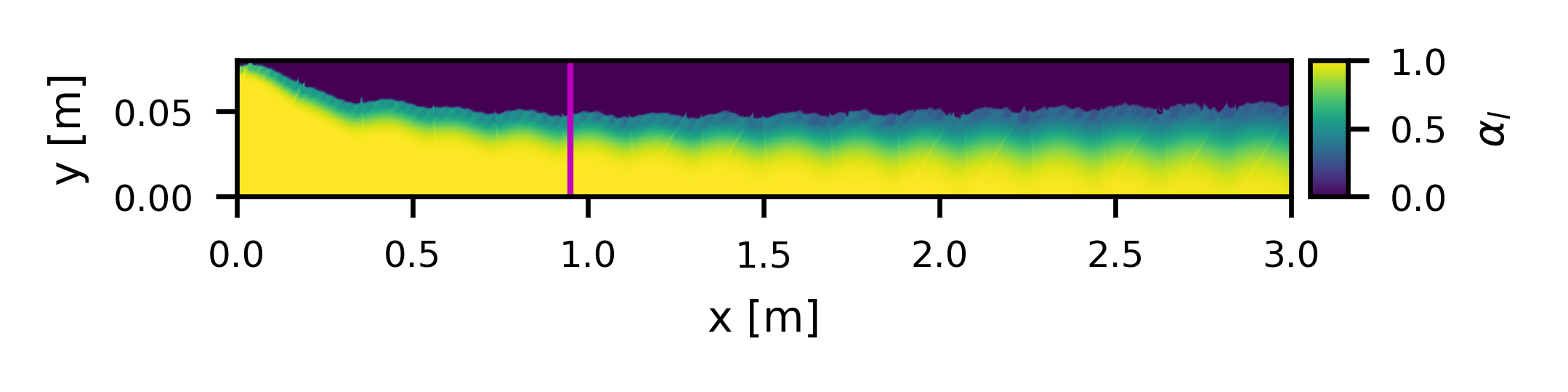}}
    \\
    \includegraphics[scale=1,trim={0.28cm 8 0.2cm
        0},clip]{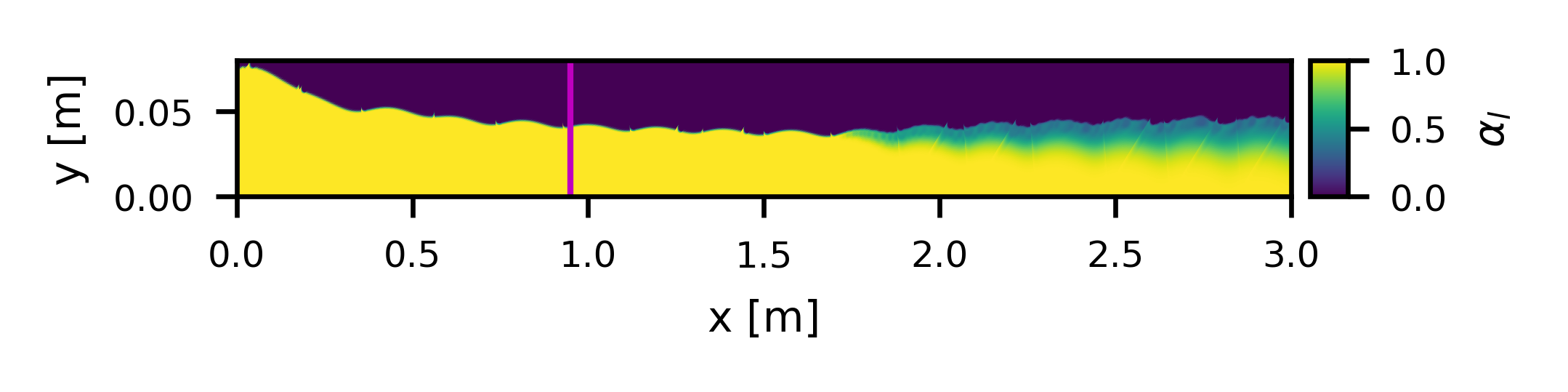}
    \caption{\textsf{F\textsubscript{s}}=4.6 -- realisable $k$-$\eps$}
    \label{subfig:icpt_fs46}
  \end{subfigure}
  \begin{subfigure}[b]{0.49\textwidth}
    \includegraphics[scale=1,trim={0.6cm 29 0.2cm
        0},clip]{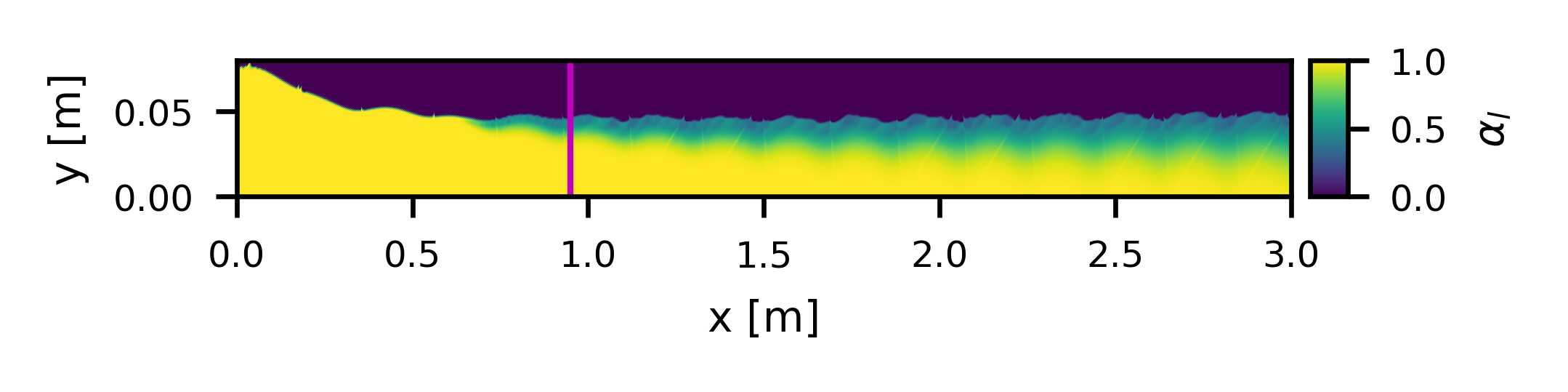}
    \\
    \includegraphics[scale=1,trim={0.6cm 29 0.2cm
        0},clip]{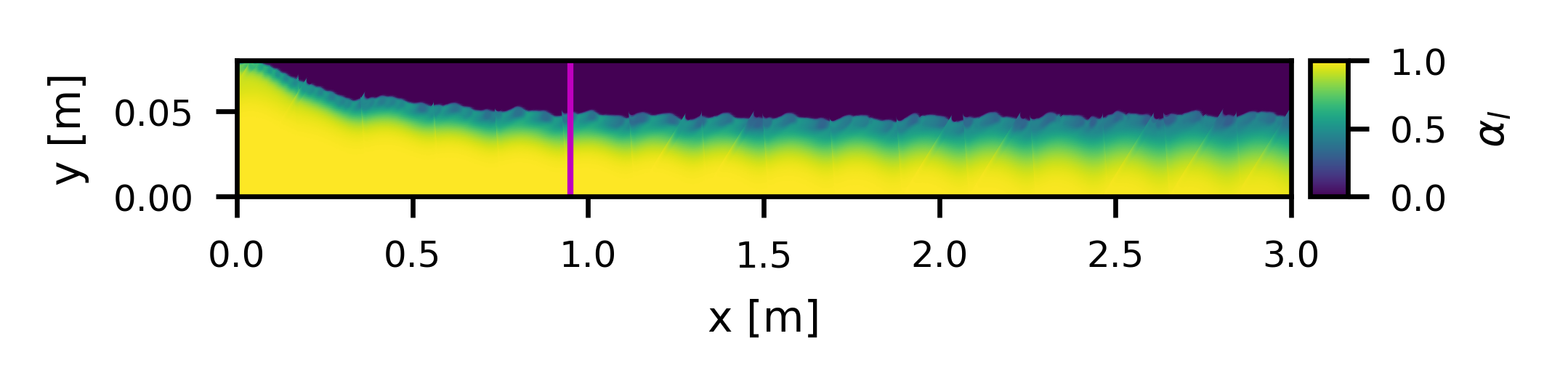}
    \\
    \includegraphics[scale=1,trim={0.6cm 8 0.2cm
          0},clip]{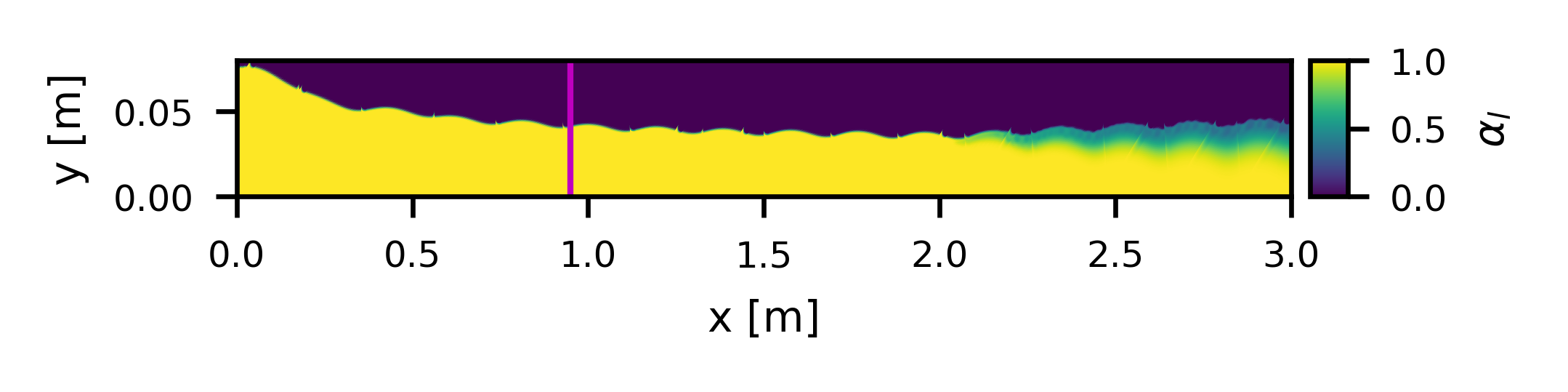}
    \caption{\textsf{F\textsubscript{s}}=4.6 -- $k$-$\omega$ SST}
    \label{subfig:icpt_fs46}
  \end{subfigure}
  \vspace{-3pt}
  \begin{subfigure}[b]{0.49\textwidth}
    \includegraphics[scale=1,trim={0.28cm 29 0.2cm
        0},clip]{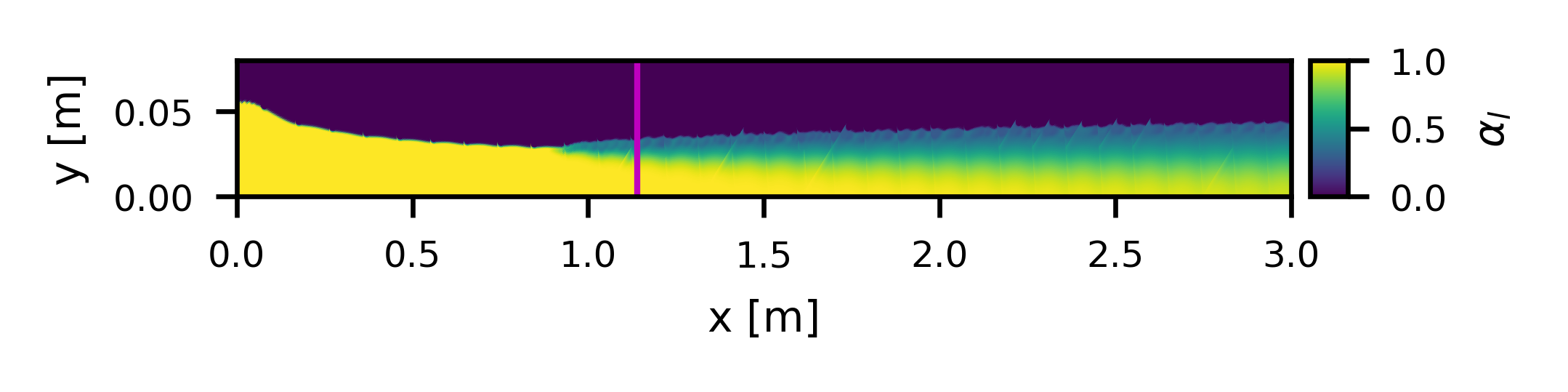}
    \\
    \includegraphics[scale=1,trim={0.28cm 29 0.2cm
        0},clip]{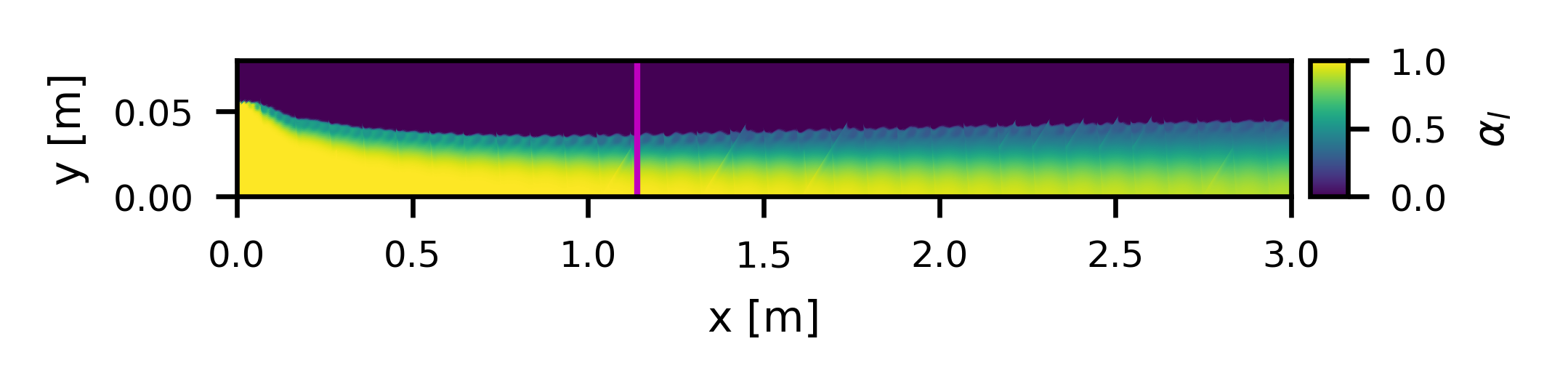}
    \\
    \includegraphics[scale=1,trim={0.28cm 8 0.2cm
        0},clip]{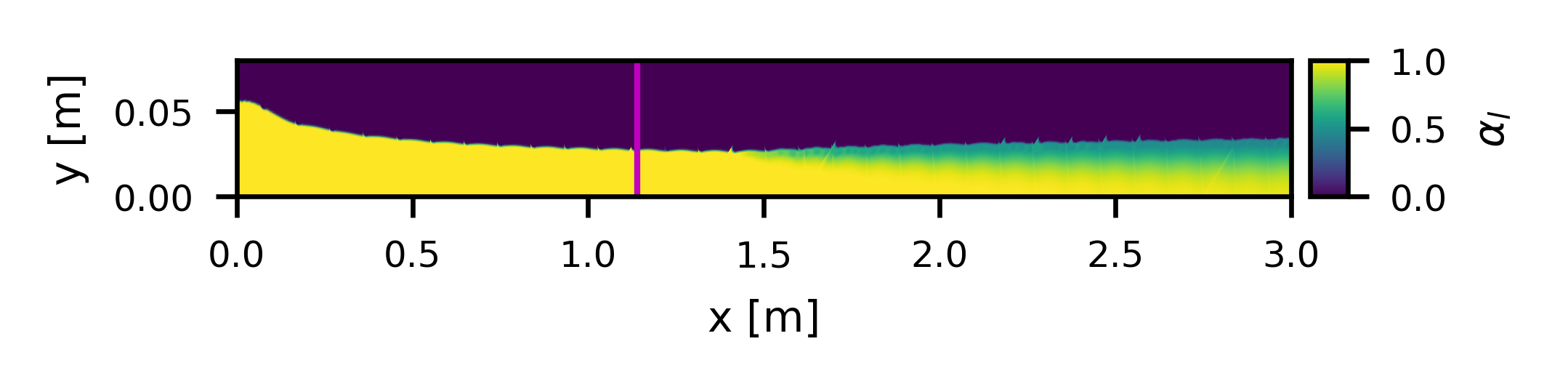}
    \caption{\textsf{F\textsubscript{s}}=8.3 -- realisable $k$-$\eps$}
  \end{subfigure}
  \begin{subfigure}[b]{0.49\textwidth}
    \includegraphics[scale=1,trim={0.6cm 29 0.2cm
        0},clip]{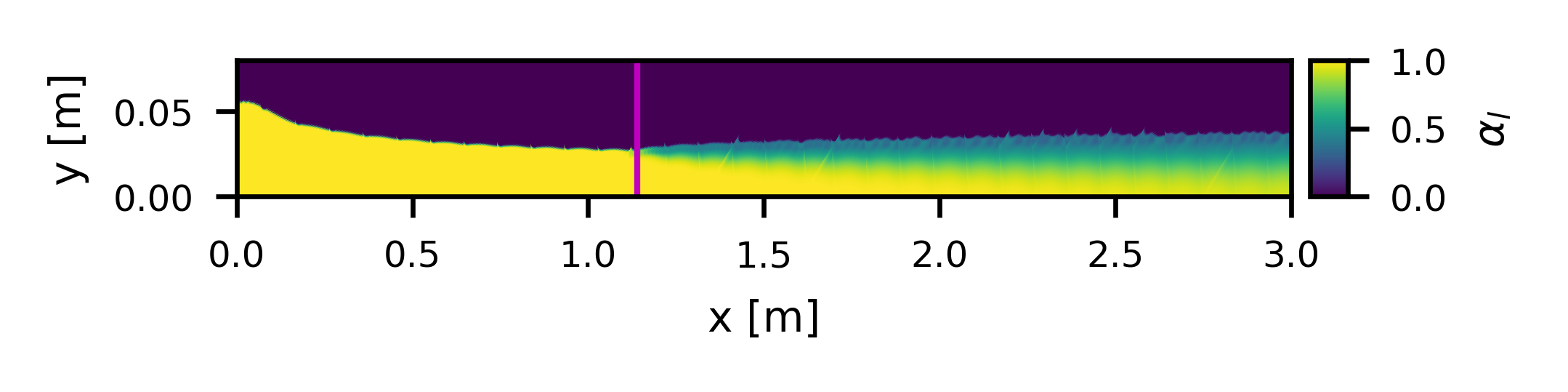}
    \\
    \includegraphics[scale=1,trim={0.6cm 29 0.2cm
        0},clip]{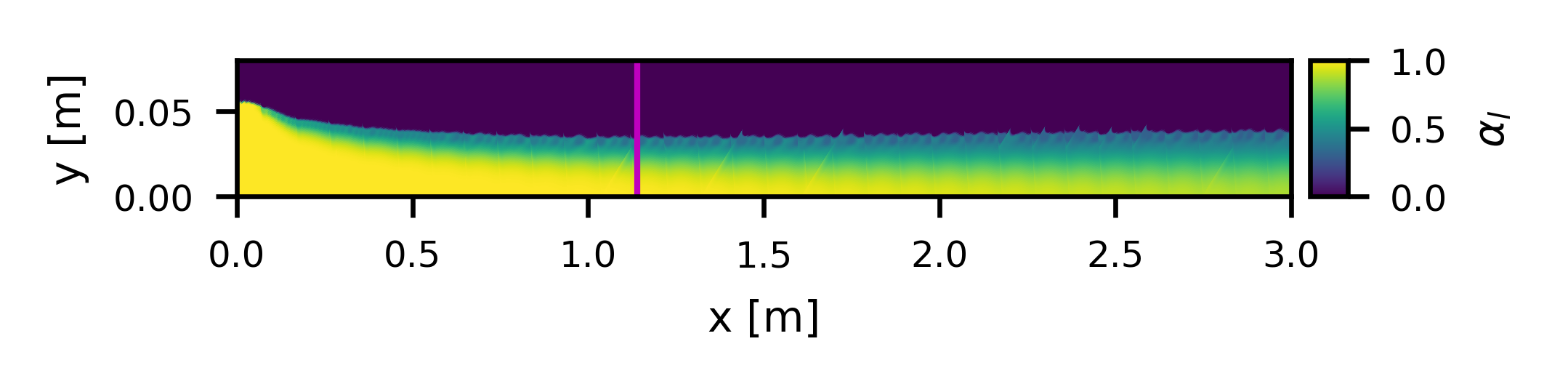}
    \\
    \includegraphics[scale=1,trim={0.6cm 8 0.2cm
        0},clip]{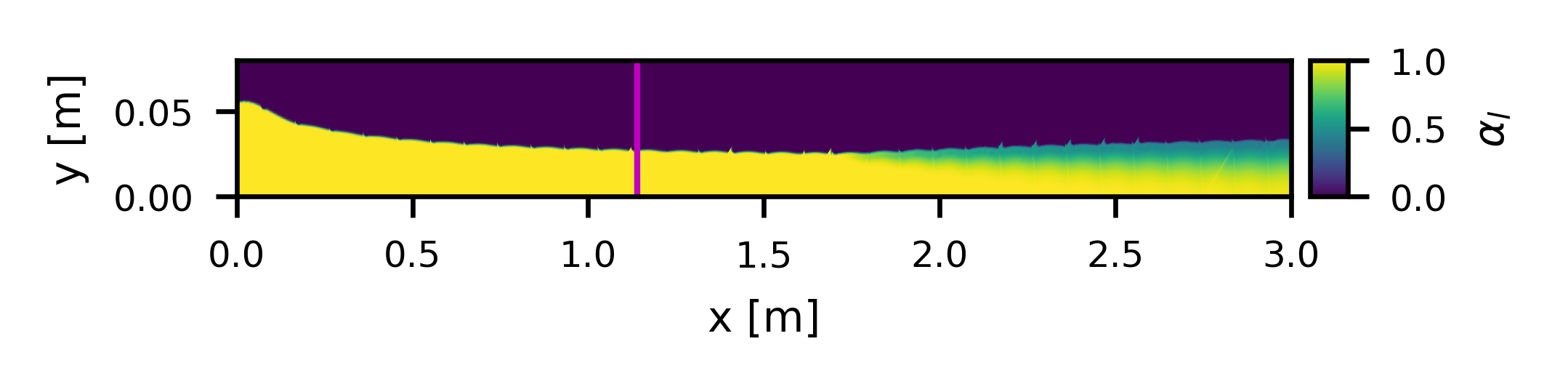}
        \caption{\textsf{F\textsubscript{s}}=8.3 -- $k$-$\omega$ SST}
        \label{subfig:icpt_fs82}
  \end{subfigure}
  \vspace{-3pt}
  \begin{subfigure}[b]{0.49\textwidth}
    \includegraphics[scale=1,trim={0.28cm 29 0.2cm
        0},clip]{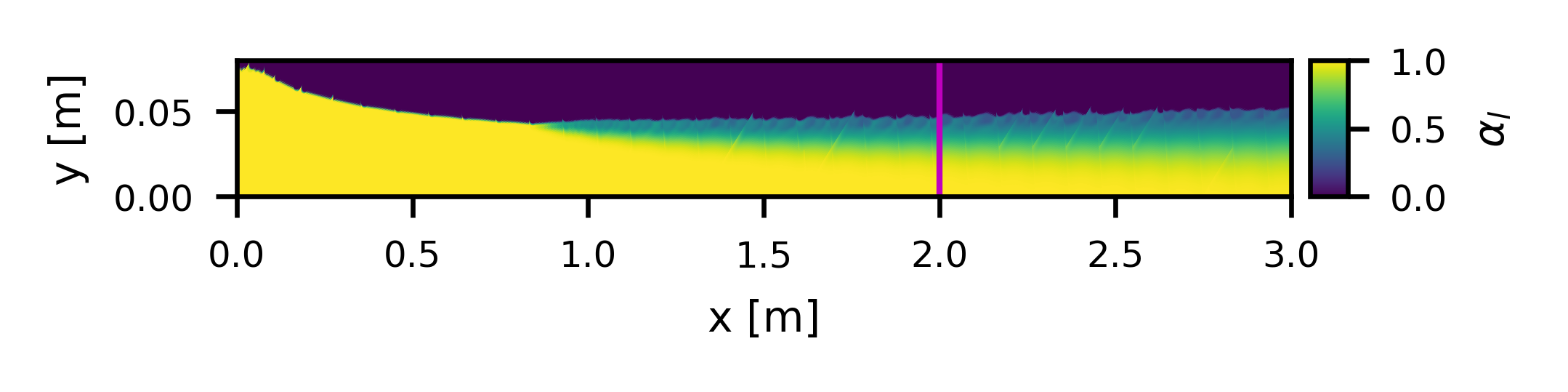}
    \\
    \includegraphics[scale=1,trim={0.28cm 29 0.2cm
        0},clip]{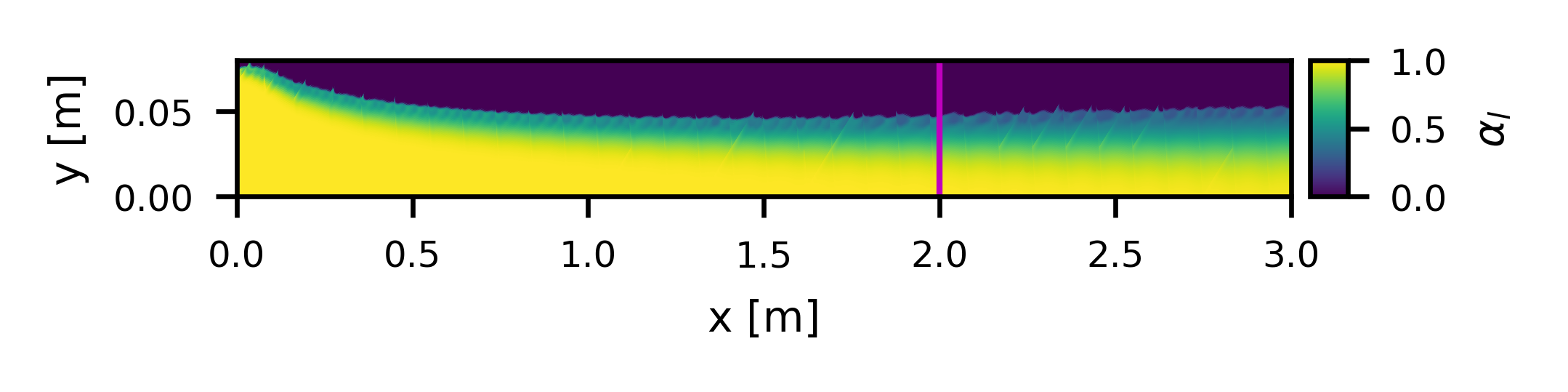}
    \\
    \includegraphics[scale=1,trim={0.28cm 8 0.2cm
        0},clip]{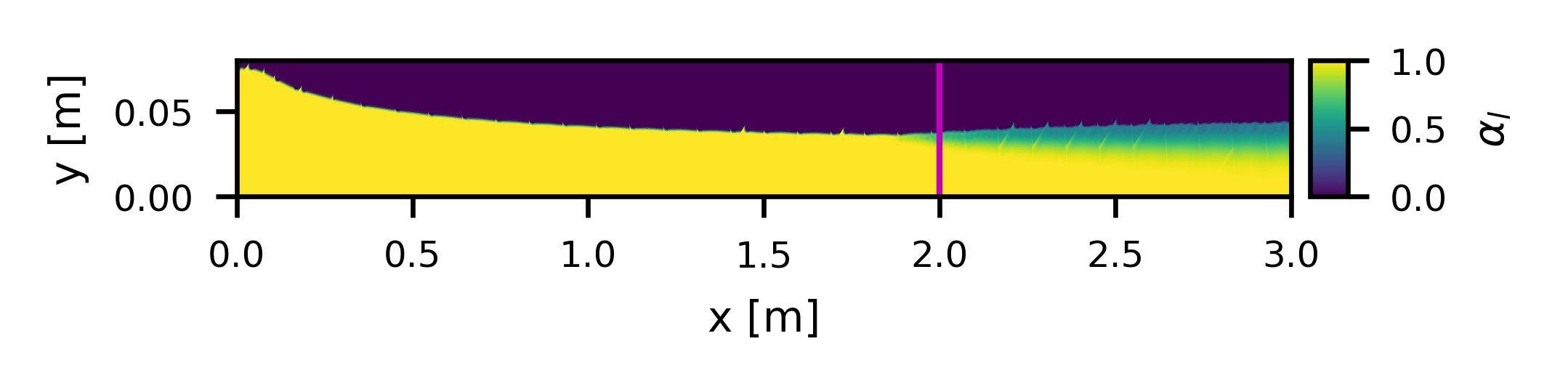}
    \caption{\textsf{F\textsubscript{s}}=13 -- realisable $k$-$\eps$}
    \end{subfigure}
  \begin{subfigure}[b]{0.49\textwidth}
    \includegraphics[scale=1,trim={0.6cm 29 0.2cm
        0},clip]{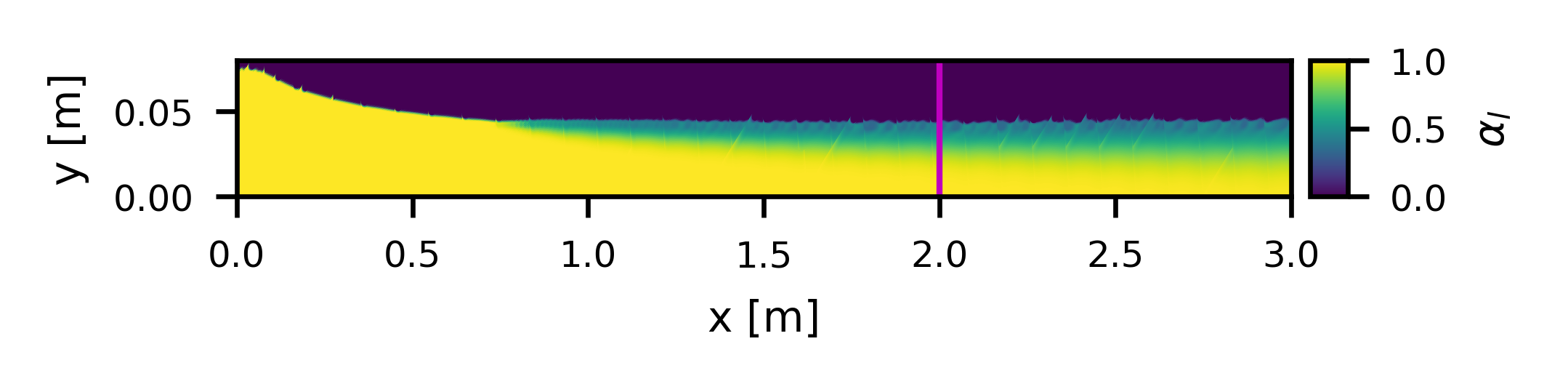}
    \\
    \includegraphics[scale=1,trim={0.6cm 29 0.2cm
        0},clip]{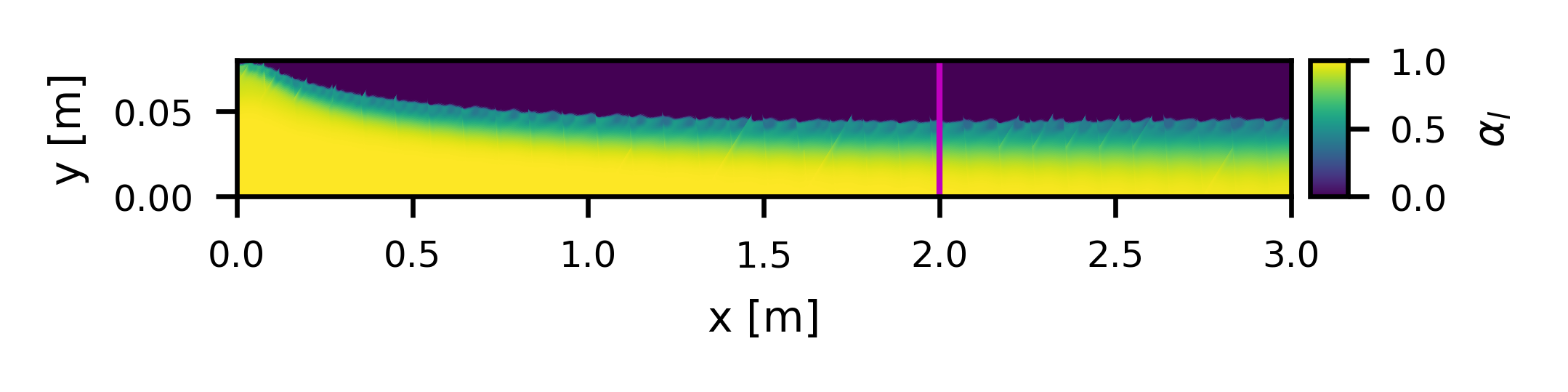}
    \\
    \includegraphics[scale=1,trim={0.6cm 8 0.2cm
          0},clip]{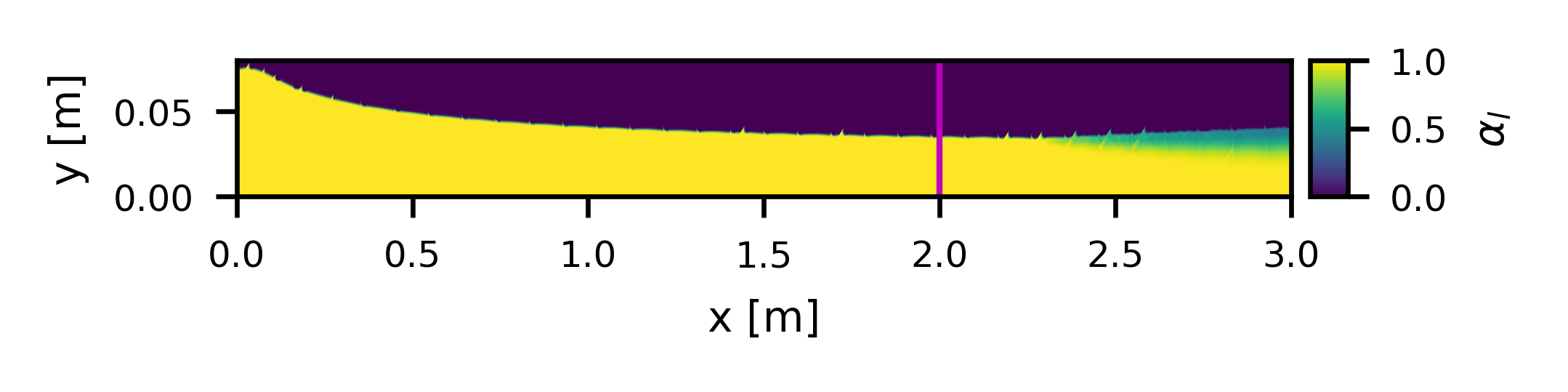}
    \caption{\textsf{F\textsubscript{s}}=13 -- $k$-$\omega$ SST}
    \label{subfig:icpt_fs13}
  \end{subfigure}
    \caption{$\alpha_l$-fields (starting at the
      psuedo-bottom) illustrating the inception point locations
      predicted using Eq.~\eqref{eq:pt}-\eqref{eq:pd} for the different
      \textsf{F\textsubscript{s}} cases on grid G3. The middle sub-figures show
      simulations using standard turbulence modelling, whilst in the lower
      sub-figures the variable density formulation is used. The top sub-figures
      refers to results using standard turbulence modelling and $k_c=0.2$ m$^2/$s$^2$, as
      suggested in \cite{Lopes2017}. The vertical lines indicate the
      experimental inception points. 
    }
  	\label{fig:spf_icpt_inv}
\end{figure}

The results are summarized in Table~\ref{tab:turb}, and the corresponding distributions of $\alpha_{air}$ are shown in Figure~\ref{fig:spf_icpt_inv}.
For the two models from the standard library, the inception of entrainment is
triggered immediately (downstream the crest) for all the considered Froude numbers.
When the \texttt{varRho} variants are employed instead, the inception point shifts downstream.
This reflects the fact that these model predict significantly lower values of $k$.
For $k$-$\omega$ SST the agreement with experimental data is nevertheless poor, but
for the realisable $k$-$\eps$ model the results are more promising.

In the results produced using standard turbulence and $k_c=0.2$ m$^2/$s$^2$, the 
inception point is predicted in relatively good agreement with the experimental
results for all cases but the \textsf{F\textsubscript{s}}=13 case, where the
correspondence is rather bad. This differs from the results attained for AIF at
the G1 grid (see Section~\ref{sec:aiffscomp}), where the inception points were poorly
predicted for both \textsf{F\textsubscript{s}}=8.3 and
\textsf{F\textsubscript{s}}=13. 
Compared to the $k_c=0.2$~m$^2/$s$^2$ criterion, the automatic activation criterion performs on
par using the realisable $k$-$\eps$ turbulence model and the variable density
turbulence framework.   

Overall, none of the tested models perform well enough to use the proposed source term activation criterion.
The improved performance of the \texttt{varRho} models shows the importance of properly accounting for the density gradient in the transport equations for the modelled flow quantities.
Improving turbulence modelling accuracy near interfaces is an active area of research.
The results presented here warrant a deeper investigation of what models are appropriate for the stepped-spillway flow.

\begin{table}[htp]
  \begin{minipage}{.64\textwidth}
  \caption{Inception points found using the source term activation
    criteria given in Eq.~\eqref{eq:pt}-\eqref{eq:pd} for SPF simulations on
    grid G3 using different models for turbulence modelling, 
    compared to physical model results by \citet{Bung2011}.}
  \label{tab:turb}
  \begin{tabular}{lllrlrr} 
    \textbf{\textsf{F\textsubscript{s}}} & \textbf{Grid} & \textbf{Turbulence} &
    \textbf{$L_{i,sim}$} & \textbf{$L_{i,expr}$} & $\Delta L_{i}$  & $\Delta n_{i}$\\
    \hline
    2.7  & G3 & $k$-$\omega$ SST, $k_c=0.2$          & 0.60 & 0.67 & -0.07 & -0.5 \\ 
    2.7  & G3 & realisable $k$-$\eps$, $k_c=0.2$     & 0.60 & 0.67 & -0.07 & -0.5 \\ 
    2.7  & G3 & $k$-$\omega$ SST              & 0.00    & 0.67 & -0.67 & -5.0  \\ 
    2.7  & G3 & realisable $k$-$\eps$                    & 0.00    & 0.67 & -0.67 & -5.0 \\ 
    2.7  & G3 & \ttt{varRho}/$k$-$\omega$ SST & 1.34 & 0.67 & 0.67  &  5.0 \\ 
    2.7  & G3 & \ttt{varRho}/realisable $k$-$\eps$       & 1.10  & 0.67 & 0.43  &  3.5 \\ 
    \hline
    4.6  & G3 & $k$-$\omega$ SST, $k_c=0.2$       & 0.57 & 0.95 & -0.38 & -2.0   \\ 
    4.6  & G3 & realisable $k$-$\eps$, $k_c=0.2$  & 0.67 & 0.95 & -0.28 & -1.5 \\ 
    4.6  & G3 & $k$-$\omega$ SST              & 0.00 & 0.95 & -0.95  & -5.0 \\ 
    4.6  & G3 & realisable $k$-$\eps$     & 0.00 & 0.95 & -0.95  & -5.0 \\ 
    4.6  & G3 & \ttt{varRho}/$k$-$\omega$ SST & 2.00 & 0.95 & 1.05  & 5.5 \\ 
    4.6  & G3 & \ttt{varRho}/realisable $k$-$\eps$  & 1.70 & 0.95 & 0.75  & 4.0 \\ 
    \hline
    8.3  & G3 & $k$-$\omega$ SST, $k_c=0.2$        & 1.05 & 1.14 & -0.09 & -0.9  \\ 
    8.3  & G3 & realisable $k$-$\eps$, $k_c=0.2$   & 0.86 & 1.14 & -0.28 & -2.9  \\ 
    8.3  & G3 & $k$-$\omega$ SST                   & 0.00 & 1.14 & -1.14  & -12.0  \\ 
    8.3  & G3 & realisable $k$-$\eps$              & 0.00 & 1.14 & -1.14  & -12.0  \\ 
    8.3  & G3 & \ttt{varRho}/$k$-$\omega$ SST      & 1.70 & 1.14 & 0.56   & -5.9   \\ 
    8.3  & G3 & \ttt{varRho}/realisable $k$-$\eps$ & 1.43 & 1.14 & 0.29  &  3.1   \\ 
    \hline
    13  & G3 & $k$-$\omega$ SST, $k_c=0.2$      & 0.76 & 2.00 & -1.24 & -13.0 \\ 
    13  & G3 & realisable $k$-$\eps$, $k_c=0.2$ & 0.86 & 2.00 & -1.14 & -12.0 \\ 
    13  & G3 & $k$-$\omega$ SST                 & 0.00 & 2.00 & -2.00  & -21.0 \\ 
    13  & G3 & realisable $k$-$\eps$            & 0.00 & 2.00 & -2.00  & -21.0 \\ 
    13  & G3 & \ttt{varRho}/$k$-$\omega$ SST    & 2.28 & 2.00 &  0.28  & 2.9 \\ 
    13  & G3 & \ttt{varRho}/realisable $k$-$\eps$  & 1.90 & 2.00 & -0.10 & -1.1 \\ 
    \hline

  \end{tabular}
  \end{minipage}
\end{table}

\section{Conclusions}
\label{sec:conclusion}

This study presents developments in numerical modelling of self-aeration in stepped spillways.
The model of~\citet{Lopes2017} is taken as baseline, and a large simulation campaign is conducted in order to explore its properties: robustness with respect to flow conditions, grid resolution, as well as sensitivity to the model parameters.
The simulations were performed for spillway geometries and inflow discharge values used in the experiments of~\citet{Bung2011}, and cover four step Froude numbers in the range from $2.7$ to 13.
The corresponding experimental data was used as reference.

The results showed that for the case of spillway flows, three of the model parameters could be removed without loss of generality, making simulation setup easier.
The main weakness of the model is shown to be its significant sensitivity to the density of the grid.
In particular, with increased resolution, the effect of the model diminishes until it is, essentially, no longer active.
Nevertheless, at selected grid resolutions, the demonstrated accuracy of the model was acceptable for all considered step Froude numbers.
Interestingly, the prediction of the mean velocity profiles was shown to not be affected by air entrainement modelling, and good results could be achieved using only the underlying VoF solver.

The main reason behind the model's deactivation on dense grids has been
identified to be the form of $\delta_{fs}$, which is the function used for
limiting the activation region of the volumetric air entrainment source term,
see Eq.~\eqref{eq:aifdeltafs}.
The region of non-zero values of $\delta_{fs}$ shrinks as the grid gets refined, and, in the limit, the source term is set to zero in the whole domain, regardless of flow conditions.
To address this issue, a new formulation for $\delta_{fs}$ is proposed, combining a parabolic profile with distance-based cut-off.
Simulations reveal that while fundamentally the results still depend on the grid resolution in the same manner, under the new definition of $\delta_{fs}$ the robustness of the model is improved.

As an additional modification, amplifying the diffusion term in the $\alpha_g$-equation~\eqref{eq:alphag} is proposed in order to account for the propagation of entrained air into corners of the steps.
Results reveal that this leads to a significant improvement in predictive accuracy, the new model performing better than the original~\citep{Lopes2017} across the whole considered range of \textsf{F\textsubscript{s}} numbers.
Furthermore, the robustness of the model with respect to grid resolution improves significantly as well.
It should be acknowledged that the selection of the value of the diffusion coefficient, $C_t = 150$, is currently not physically motivated and can, therefore, be called into question.
Nevertheless, we believe that the possibility to use the same value across different flow conditions and the clearly demonstrated advantages in terms of the performance of the model are sufficiently strong arguments in favour of adopting the proposed modification.
Finding a more rigorous connection between~$C_t$ and the characteristic scales of the flow remains as a line of future work.

Finally, an algorithm for automatic estimation of the inception point is tested.
The criterion for the inception point is based on energy balance, as proposed by~\cite{hirt2003modeling}.
The performance of the algorithm are heavily dependant on the underlying turbulence model.
Unfortunately, for the four considered models the predictions were not reliable, highlighting the need for a more careful investigation of what turbulence modelling is appropriate for self-aerating multiphase flows.

\section*{Acknowledgments}
This work was supported by grant number P38284-2 from the Swedish Energy
Agency. The simulations were performed on resources provided by Chalmers Centre
for Computational Science and Engineering (C3SE), the NTNU IDUN/EPIC computing
cluster, and UNINETT Sigma2 -- the National Infrastructure for High Performance
Computing and Data Storage in Norway. The authors are thankful to Daniel Bung
for sharing the data files from his physical experiments and to Pedro Lopes for sharing the source code of the \texttt{airInterFoam} solver.

\newpage
\bibliographystyle{elsarticle-harv} 
\bibliography{main}

\end{document}